\documentclass[%
 reprint,
 amsmath,amssymb,
 aps,
]{revtex4-2}

\usepackage{graphicx}
\usepackage{dcolumn}
\usepackage{bm}
\usepackage{xcolor}
\usepackage{subfiles}
\usepackage{wrapfig}
\usepackage{xr}
\newcommand{\SK}[1]{\textcolor{black}{{#1}}}

\newcommand{\JH}[1]{\textcolor{black}{{#1}}}

\begin{document}

\preprint{APS/123-QED}

\title{Active Particle Doping Suppresses Brittle Failure in Ultrastable Glasses}

\author{Rashmi Priya$^1$} 
\author{J\"urgen Horbach$^2$}
\author{Smarajit Karmakar$^1$} 
 \email{smarajit@tifrh.res.in}
\affiliation{ $^1$Tata Institute of Fundamental Research, 36/P, Gopanpally Village, Serilingampally Mandal,
Ranga Reddy District, Hyderabad, India 500046\\ $^2$Institut für Theoretische Physik II - Soft Matter, Heinrich-Heine-Universität, Düsseldorf, Germany}

\begin{abstract}
\JH{Ultrastable glasses are known for their exceptional mechanical stability but often fail in a brittle manner, typically marked by the formation of shear bands when subjected to shear deformation. An open question is how shear banding is affected by active particles. Here, we address this issue by investigating ultrastable glasses that are doped with self-propelled particles (SPPs) that perform run-and-tumble motions. In the presence of active particles we find a crossover from heterogeneous to homogeneous yielding. Through extensive computer simulations of a polydisperse model, which is capable of producing ultrastable glasses using swap Monte Carlo Methods, we demonstrate the progressive emergence of multiple shear bands under activity, leading to continuous and delayed yielding. Interestingly, we uncover a compensatory relationship between active forces and global shear rates: a rise in one can offset a decline in the other, arising from the isomorphic-like behavior exhibited by different combinations of active forces and strain rates.  We also identify a non-monotonic relationship between yielding and persistence time: while the yield stress increases at shorter persistence times with increasing active forces, it tends to decrease with longer persistence times. This observation highlights a tunable range of activity that can modify the yielding mode. Lastly, we show that the spacing between shear bands diminishes in a power law manner as the magnitude of the active force increases.}

\end{abstract}

\maketitle


\noindent{\label{sec:level1}\bf\large Introduction:}
\SK{Ultrastable glasses achieve high density and reduced enthalpy compared to conventional glasses, leading to exceptional mechanical stability \cite{2007Sci...315..353S, 2022NCimR..45..325R, doi:10.1126/sciadv.aau5423}. Achieving such high mechanical stability through bulk cooling necessitates an extremely slow cooling rate, typically ranging from $10^5$ to $10^{10}$ times slower than the normal cooling rate required for glass formation \cite{Arceri2022}. As a consequence, the preparation of such glasses {\it in silico} and {\it in vitro} has been exceptionally challenging. Swap Monte-Carlo has been an efficient technique for investigating these systems through computer simulations, leveraging particle size polydispersities for swaps and circumventing such difficulties \cite{PhysRevE.63.045102, PhysRevX.7.021039, PhysRevLett.125.085505, Berthier_2019}. However, the feasibility of implementing this method in experiments to obtain ultrastable glasses remains unexplored. Recently, a groundbreaking method introduced in Ref.~\cite{doi:10.1126/science.1135795, 10.1063/1.5006265} proposed a vapor deposition technique for obtaining thin film samples, primarily relying on surface deposition. Despite encountering some geometrical constraints, this approach marked a significant advancement. Due to a recent surge in interest in understanding failure mechanisms in well-annealed glasses relevant to experiments and real-life applications, many proposals are made to generate and study the behavior of the ultrastable glasses in laboratories and computer simulations, such as by replica-exchange MC or shoving algorithm swap Monte-Carlo, by minimizing augmented potential energy, random particle bonding, and introducing point-like or rod-like inclusions \cite{1992EL.....19..451M,1996JPSJ...65.1604H, PhysRevLett.123.185501, Ozawa2023, doi:10.1073/pnas.1222848110, mutneja2023yielding}, etc.}

\SK{Ultrastable glasses have} \JH{the potential to be used in} \SK{many practical and industrial applications due to their high thermodynamic, kinetic, and mechanical stability and are the subject of extensive research. When these mechanically stable glasses reach their yield stress, they fail abruptly, characterized by a macroscopic stress drop and associated by stress localization along a plane as a shear band. These system-spanning shear bands are the hallmark of brittle failure in ultrastable glasses.}
\JH{Various studies have been conducted on how yielding and shear band formation are affected by modifications of the internal properties of glass states \cite{PhysRevE.57.7192, greer2013shear, hufnagel2000development, sergueeva2005shear, maass2015shear, sun2012serrated, PhysRevResearch.2.023203, RevModPhys.90.045006}. These include incorporating point-like or rod-like inclusions, examining random particle bonding, and characterizing delayed yielding and stress overshoot \cite{PhysRevLett.109.255502, PhysRevMaterials.6.065602, PhysRevLett.123.185501, mutneja2023yielding}. However, there is a scarcity of studies on how active particles affect the brittleness of glasses. Here, we show that in ultrastable glasses active particles destabilize the shear localization to a narrow band. Thus, active particles may be used as local shearing agents in ultrastable glasses to systematically circumvent their brittle mechanical response during yielding.}

\SK{Active systems are fascinating and complex, as they are inherently out of equilibrium in nature, and they do not follow detailed balance, being driven either internally or by external forcing \cite{PhysRevLett.75.1226, PhysRevE.58.4828, ramaswamy2010mechanics, doi:10.1126/science.1230020, fily2012athermal, henkes2011active}. Active matter is a system in which the constituents can move internally, driven by their internal energy, in addition to the environmental influence of thermal fluctuations. Active systems exhibit a wide range of interesting dynamical phenomena, including spontaneous symmetry breaking of the rotational order in two dimensions, leading to the formation of ordered phases of clusters or flocks \cite{PhysRevLett.75.1226, PhysRevE.58.4828}. These clusters have coherent collective motion at low noise strength and high particle density. Many biological systems exhibit collective dynamical behavior, in which forces generated by ATP consumption drive the dynamics instead of thermal fluctuations. A simple model of these systems that can capture some of the salient dynamical behaviors is a collection of self-propelled particles (SPPs) \cite{RevModPhys.85.1143}. Studies show that collective dynamics of cells and tissues during cell proliferation, cancerous cell progression, and wound healing \cite{doi:10.1073/pnas.0901462106, doi:10.1073/pnas.1010059108, PARRY2014183, Park2015, doi:10.1073/pnas.1510973112, Prosseda2010-la, Nishizawa2017, doi:10.1126/science.1135795, Vishwakarma2020-ak} have dynamical features similar to glassy dynamics. These results have led to the emergence of a new research direction on disordered systems called active glasses \cite{mandal2016active, mandal2021study, paul2023dynamical, Mandal2020}. Active glasses refer to a class of materials consisting of self-propelled particles (SPPs) that exhibit glass-like behavior characterized by slow and heterogeneous dynamics.}

\JH{Recent studies \cite{doi:10.1073/pnas.2019909118, doi:10.1073/pnas.2318917121} have emphasized similarities between the effect of shear and activity. For a system of run-and-tumble active particles, Sharma {\it et al.}~\cite{Sharma2025, Goswami_2025_YieldingActive} have demonstrated similarities to oscillatory shear with regard to annealing and yielding.} Studies have also shown that that the brittle-to-ductile transition in dense colloidal gels can be controlled through active doping \cite{Zhou_2024_ActiveDoping}. \JH{As a common feature, activity tends to fluidize amorphous solid states by breaking cages above a critical active force $f_0$ for a given persistence time $\tau_p$ \cite{mandal2016active}. Analogously, there is fluidization driven by activity via cellular metabolism in bacterial cytoplasm \cite{PARRY2014183}. In this work, however, we find a less intuitive dependence on persistence time: at low $\tau_p$, activity shows non-temperature-like behavior, while at higher $\tau_p$ fluidization as described above is observed. By exploring similarities between activity and shear, as in previous studies \cite{doi:10.1073/pnas.2019909118, doi:10.1073/pnas.2318917121, Sharma2025, mandal2016active, PARRY2014183}, we elucidate the possibilities of active particles as local shearing agents both at low and high persistence times and investigate their impact on the mechanical stability of ultrastable glasses at finite temperature and shear rate. In fact, we find that through the active forces $f_0$ and the persistence times $\tau_p$, activity can be used to control flow patterns in sheared systems.}

\JH{As in the recent study by Lamp {\it et al.}~\cite{10.1063/5.0086626}, this study investigates brittle yielding and the formation of shear bands at finite shear rates and temperatures, here with the addition of self-propelled active particles. We note that an athermal approach using an athermal quasistatic limit (AQS) scheme \cite{Utz2000} would imply the choice of an infinite persistence time and is therefore not appropriate in our case since we are interested in the behavior asspciated with finite persistence times.}



\JH{We show that active forces can enhance the mechanical stability of ultrastable glasses, as reflected in a broad overshoot in the stress-strain relation that marks the onset of plastic flow when the system yields. This is unlike the original brittle response of a ultrastable glass that is associated with a sharp stress drop and the emergence of a shear band at yielding. Instead, the doping with active particles leads to a more homogeneous flow pattern. At yielding, there is the occurrence of multiple shear bands at intermediate values of $f_0$ that with further increasing $f_0$ evolves into a network of mobile bands.}

\JH{Strategies for the avoidance of brittleness and catastrophic failure have been recently developed also for other materials. For example, a recent study on a mechanical metamaterial \cite{Surjadi2025} demonstrates an increase in both stiffness and stretchability, along with substantial energy dissipation through progressive fracture under large deformation. This is achieved by distributing defects in a checkerboard pattern that can be seen as an analogue to the emerging shear band network in our system with a doping of active particles.} 

\SK{To model the system, we prepare ultrastable glasses in 2D ($N = 10000$) and 3D ($N=10000,\, 100000$) using polydisperse soft spheres, equilibrated via molecular dynamics (MD) and swap Monte Carlo (swapMC) methods. Active particles performing run-and-tumble motion are added at a fixed concentration ($c = 0.1$) and controlled by tuning persistence time ($\tau_p$) and active forces ($f_0$). The system with active particles is subjected to an external shear at a fixed finite temperature ($T$) and finite shear rates ($\dot\gamma$). A detailed description can be found in the Model and Simulation Details in the Methods section.}


\vskip 0.1in
\noindent{\bf \large Rounded Yielding and Enhanced Mechanical Stability:}
\begin{figure*}
\includegraphics[width=0.99\textwidth]{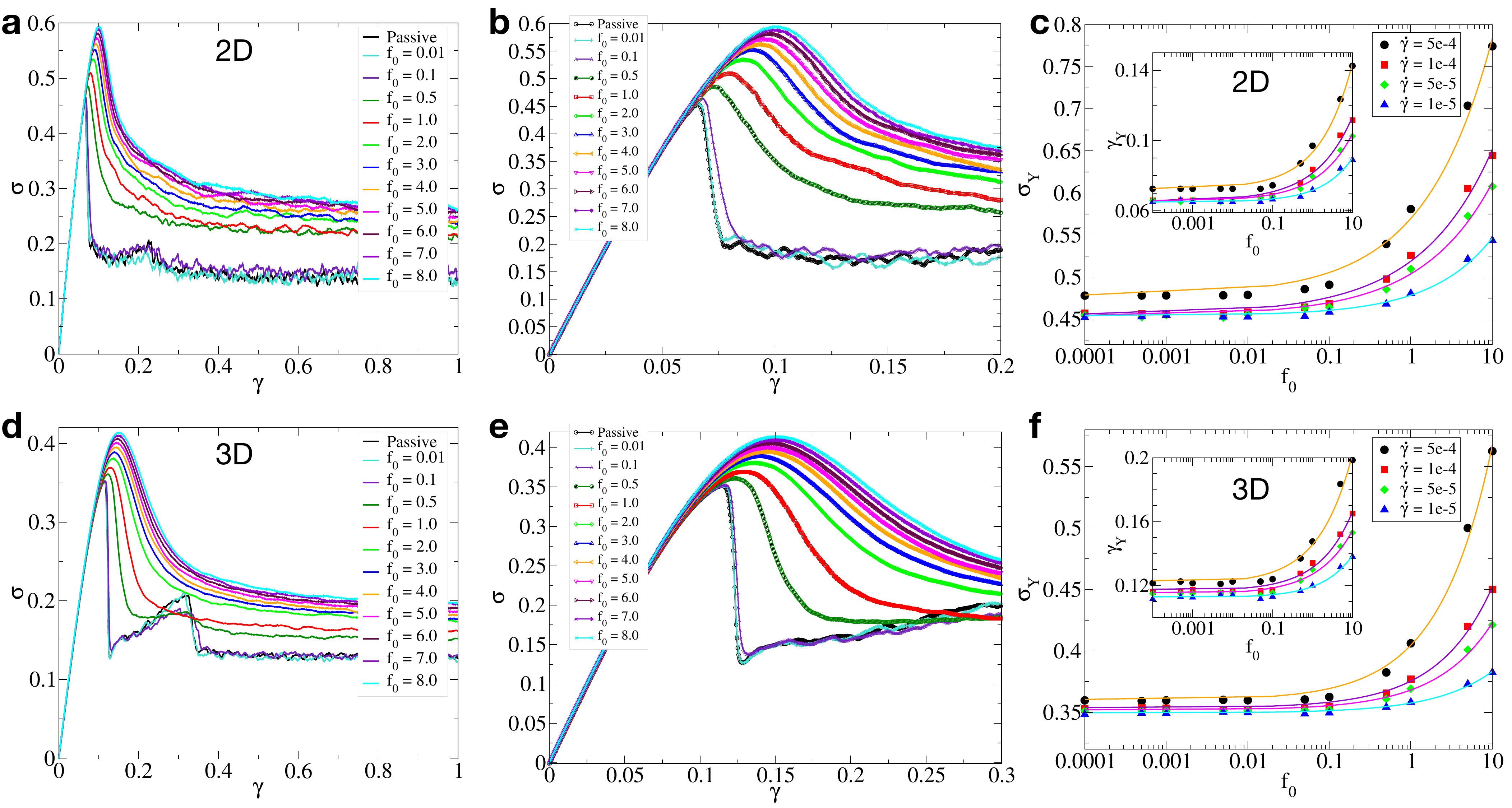}
\caption{{\bf Rounded Yielding by Activity:} Panel (a) shows the stress-strain curves with increasing $f_0$ in 2D, while panel (d) represents the curves for 3D, in both cases at a strain rate of $\dot{\gamma} = 5\times 10^{-5}$. All plots correspond to a small persistence time $\tau_p = 0.1$ and an active particle concentration of $20\%$. We observe that yielding becomes more continuous with increasing $f_0$, even at low $f_0$ values. The system requires higher external driving to reach the yield point, and the subsequent stress drop is no longer abrupt as in the passive case at the same shear rate. The zoomed-in views of (a) and (d), shown in (b) and (e), respectively, clearly demonstrate that the stress drops become increasingly rounded and less sharp. This behavior is accompanied by an increase in steady-state stress as $f_0$ increases. The increase in yield stress is evident from yield stress vs.~$f_0$ plots at different strain rates ($\dot{\gamma} = 5 \times 10^{-4}, 1 \times 10^{-4}, 5 \times 10^{-5}, 1 \times 10^{-5}$), as shown in panels (c) and (f). Panel (c) illustrates $\sigma_Y$ vs $f_0$ in 2D, with an inset displaying the corresponding yield strain ($\gamma_Y$) as a function of $f_0$ for the same shear rates. Similarly, panel (f) presents $\sigma_Y$ vs.~$f_0$ in 3D, along with an inset displaying the corresponding yield strain ($\gamma_Y$) as a function of $f_0$ for the same shear rates. The yield stress and $f_0$ follow a Herschel-Bulkley behavior according to Eq.~(\ref{HBstress}) with $\beta_1 < 1$ (data shown by points and fitting shown by curves). A similar trend is observed when plotting yield strain against increasing $f_0$ at different strain rates. The increase in yield strain is fitted by Eq.~(\ref{HBstrain}) with $\beta_2 < 1$. All plots indicate an increased mechanical stability at low $\tau_p$, signified by an increase in yield stress and strain with the addition of active forces. This occurs because, at sufficiently low $\tau_p$, shear band propagation is inhibited by active particles.}
\label{Rounded_Yielding}
\end{figure*}

\begin{figure}
\includegraphics[width=0.45\textwidth]{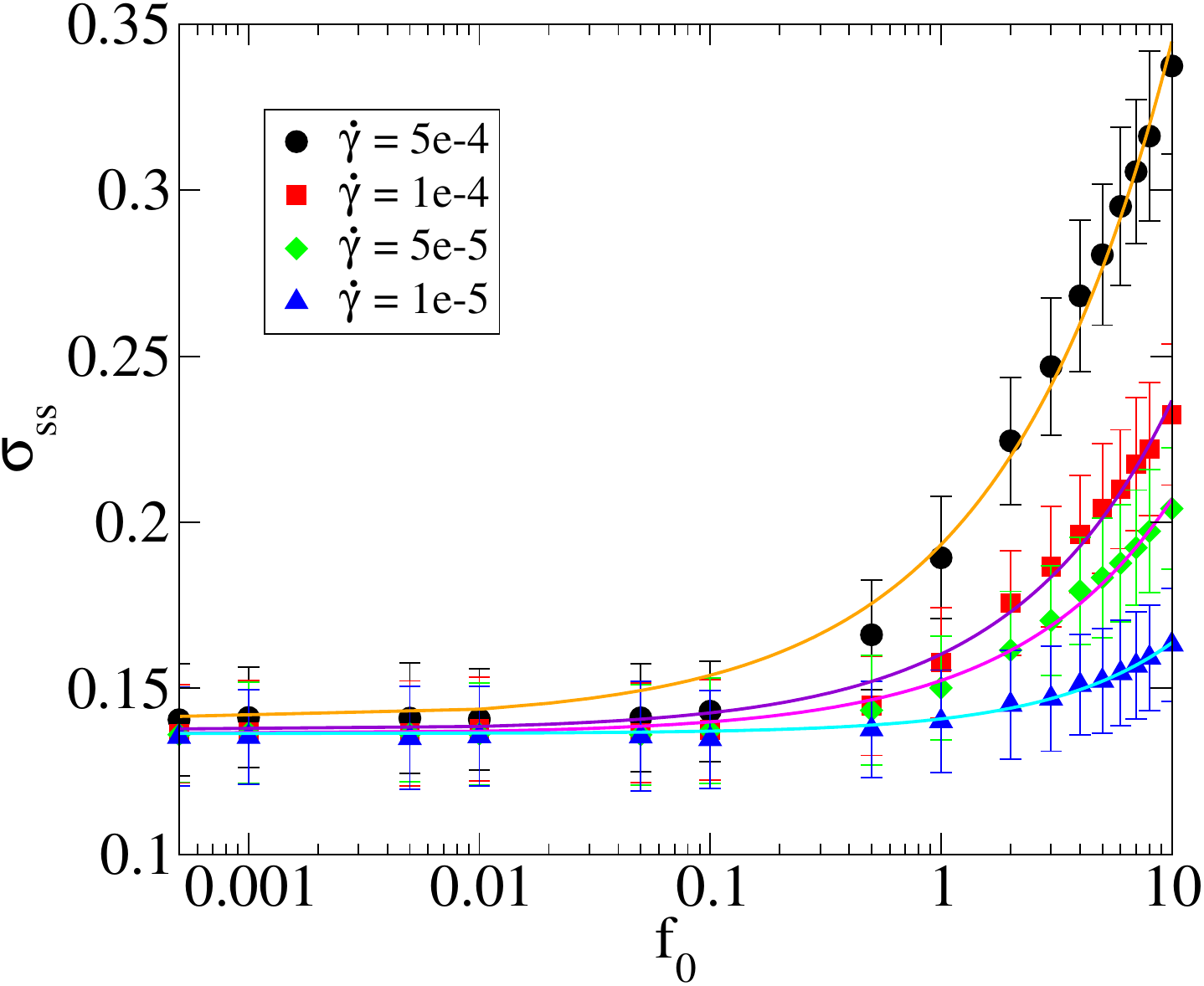}
\caption{{\bf Flow curve in 3D:} Steady-state stress for different strain rates $\dot{\gamma} = 5 \times 10^{-4}, 1 \times 10^{-4}, 5 \times 10^{-5}, 1 \times 10^{-5}$ as a function of active force $f_0$ for a system of $N=2000$ particles. The error bars represent fluctuations across different ensembles. The solid lines correspond to fits using the Herschel-Bulkley law, indicating shear-thinning behavior with a flow index $\beta_{\mathrm{s1}} < 1$ across all strain rates.}
\label{Steady-State_3D}
\end{figure}

\SK{To understand how activity influences the response of the system under shear, we plot the stress-strain curve of a passive system alongside that of an active system (Additional data for the passive system in three dimensions (3D) is shown in the Supplementary Fig.~1). Compared to the passive system, the yielding in the presence of active particles is more pronounced as the stress peak increases, but it gets rounded, as seen in Fig.~\ref{Rounded_Yielding}(a) and (d) in two (2D) and three dimensions (3D), respectively. Panels (d) and (g) show data for 3D systems for system sizes $N = 10,000$ and $100,000$, respectively, at a strain rate of $\dot{\gamma} = 5\times 10^{-5}$. These figure panels show an increase in yield stress and yield strain with increasing active force, $f_0$. The curves correspond to a system with a concentration of $c = 20\%$ active particles and a persistence time, $\tau_p = 0.1$. The behavior of increase in yield stress can be clearly observed by plotting yield stress, $\sigma_Y$ as a function of $f_0$ for various strain rates ($\dot{\gamma} = 5\times 10^{-4}, 1\times 10^{-4}, 5\times 10^{-5}, 1\times 10^{-5}$) as shown in Fig.~\ref{Rounded_Yielding}(b), (e) and (h). As the yield stress increases with increasing $f_0$ as well as considering that in the limit $f_0 \rightarrow 0$, the yield stress should reduce to the passive case, $\sigma_Y^P$,} \JH{we fitted our data using the equation}
\begin{eqnarray}
  \sigma_Y (\dot\gamma, f_0) = \sigma_Y^P(\dot\gamma) + {\alpha_1} f_0^ {\beta_1}, 
   \label{HBstress}
\end{eqnarray}
\SK{where $\alpha_1$ and $\beta_1$ are the consistency and flow indexes, which are strain rate-dependent quantities, and $\sigma_Y^P$ represents the yield stress for a passive system ($f_0 = 0$) for the respective strain rates. By fitting the yield stress data, we found that $\beta_1 < 1$ for all strain rates, indicating a Herschel-Bulkley-like shear thinning behavior with active force magnitude. This also shows that the critical value of $f_0$ required to observe an increase in mechanical yielding decreases with an increase in $\dot{\gamma}$ for $\tau_p$ $= 0.1$. Thus, one can tune the value of $f_0$ for a given $\dot{\gamma}$ at a $\tau_p$ to achieve the desired yielding. A similar yield stress curve is presented for other $\tau_p$ values $0.05, 0.5, 1.0$ in the Fig.~2 of the supplementary material. It is observed that the shear-thinning behavior diminishes as $\tau_p$ increases to $1.0$, suggesting that lower $\tau_p$ values play a unique role in exhibiting this new phenomenon. This phenomenon will be discussed further in the later part of the article. Additionally, the system requires a larger strain to reach the yield point, as seen in the rightmost column of Figs.~\ref{Rounded_Yielding}(c), (f), and (i), where yield strain is plotted as a function of $f_0$. We find a similar behavior; i.e., the yield strain increases with $f_0$ as given by
\begin{eqnarray}
   \gamma_Y(\dot\gamma, f_0) = \gamma_Y^P(\dot\gamma) + {\alpha_2} f_0^ {\beta_2},
   \label{HBstrain}
\end{eqnarray}
where the parameters $\alpha_2$ and $\beta_2$ in Eq.~\ref{HBstrain} are strain rate-dependent quantities. By fitting the yield strain data, we found that $\beta_2 < 1$ for all strain rates where $\gamma_Y^P$ represents the yield strain for each respective strain rate.}

\begin{figure*}
\centering
\includegraphics[width=0.99\textwidth]{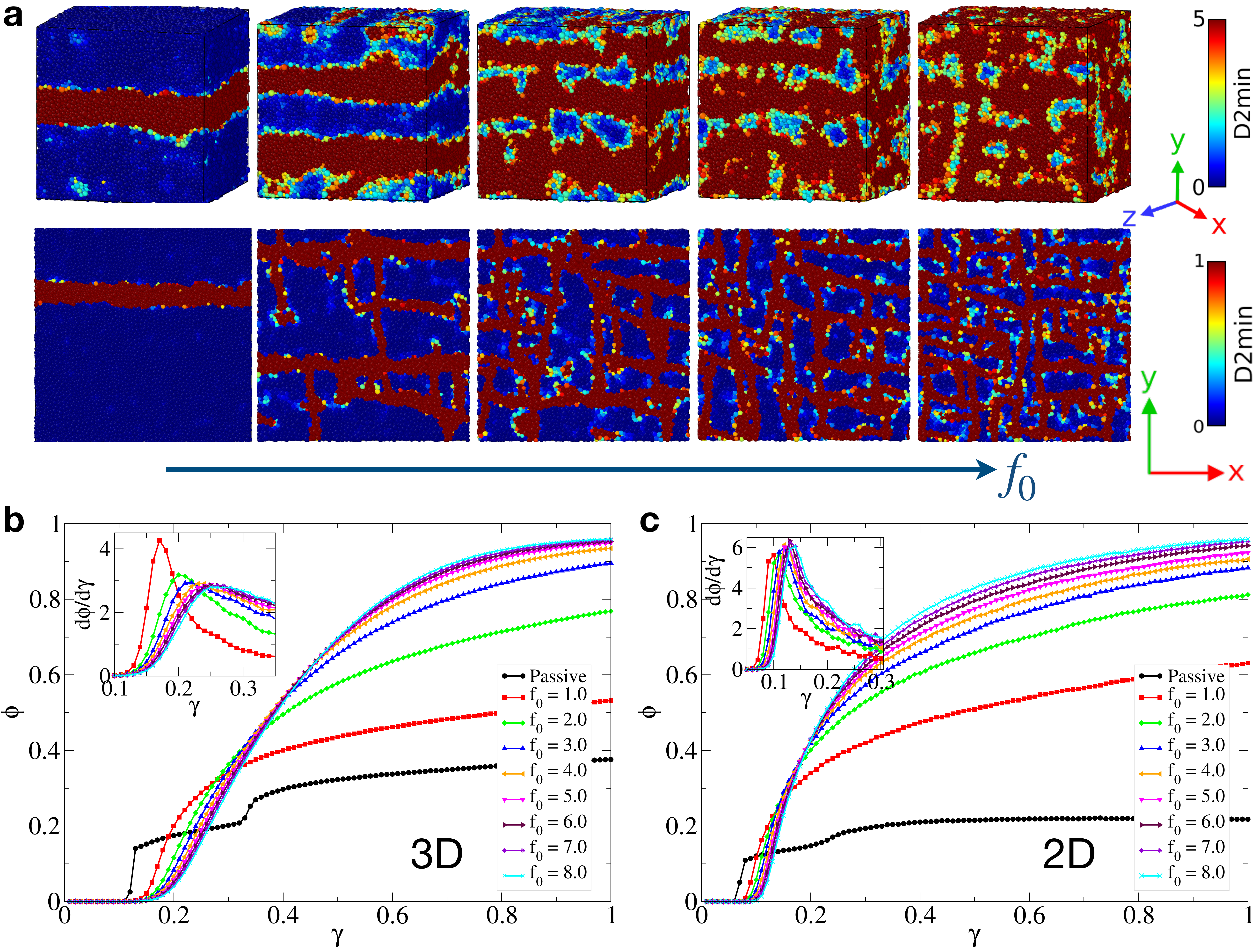}
\caption{{\bf Multiple shearbands:} The top panel of (a) shows shear bands in 3D for $N=100000$, while the bottom panel shows shear bands in 2D for $N=10000$ at a strain rate of $\dot{\gamma} = 5\times 10^{-5}$. The color bar represents the non-affine displacements calculated using $D^2_{min}$. We observe snapshots of shear bands at fixed strains of $\gamma = 0.25$ for 3D and $\gamma = 0.20$ for 2D with increasing $f_0$ ($0.0$, $1.0$, $2.0$, $4.0$, $8.0$). As $f_0$ increases, the number of shear bands also increases, which explains the behavior of the stress-strain curve shown in Fig.~\ref{Rounded_Yielding}. In (b) and (c), we have plotted the average volume fraction $\phi$ occupied by the shear band region as a function of strain ($\gamma$) in 3D and 2D systems at $\tau_p = 0.1$, respectively. The passive case is shown alongside active cases with $f_0$ values ranging from 1.0 to 8.0. The results indicate a gradual increase in the volume fraction occupied by shear bands due to the emergence of multiple shear bands. Initially, the increase is slowest for larger $f_0$, but as strain progresses, the largest $f_0$ exhibits the maximum volume fraction. This occurs because of an increase in yield strain with increasing $f_0$, which results in shear band formation at larger strains and slower progression due to gradual failure, corresponding to the more rounded behavior of the stress-strain curve.}
\label{MultipleSB}
\end{figure*}

\begin{figure*}
\includegraphics[width=0.99\textwidth]{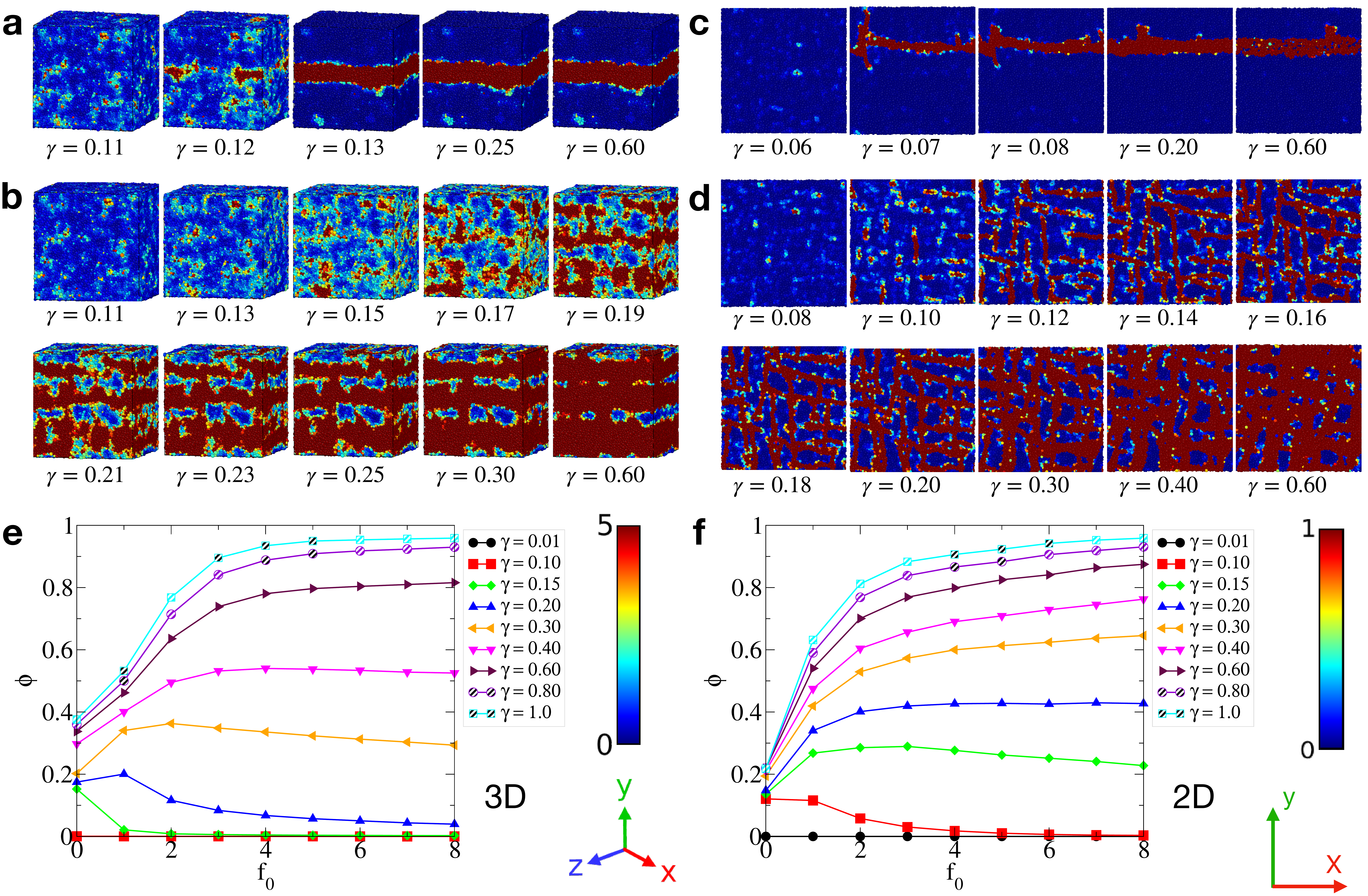}
\caption{{\bf Evolution of the shear band with increasing strain:} The evolution of shear bands in passive and active systems is shown by calculating the $D^2_{min}$ of particles under shear. The shear bands can be located using higher $D^2_{min}$ values ($1$ and $5$ in 2D and 3D)), as shown in the color bar. Panel (a) is for a 3D passive system; the formation of the shear band is abrupt, transitioning from $\gamma = 0.12$ to $\gamma = 0.13$ at a strain rate of $\dot{\gamma} = 5\times 10^{-5}$. However, in the active case (b), at $\tau_p = 0.1$ and $f_0 = 2.0$, the system forms shear bands gradually from $\gamma = 0.13$ and completes shear band formation near $\gamma = 0.25$, explaining the continuous yielding observed earlier in the stress-strain curve for the active case. Similar behavior is observed in 2D for both (c) passive ($\gamma = 0.06$ to $\gamma = 0.07$) and (d) active cases ($\gamma = 0.08$ to $\gamma = 0.20$) for $\tau_p = 0.1$ and $f_0 = 4.0$. When plotting the average volume fraction with increasing $f_0$ at different strains (e) and (f), we observe an increase in the region occupied by the shear band with increasing $f_0$. Initially, at low strain, when there is no shear band, the volume fraction is $0$ or close to $0$. As strain increases, the curve is non-zero only for the passive case at $\gamma = 0.15$ due to the formation of the shear band. With further increase in strain, we observe a non-monotonic increase in volume fraction due to the delayed yielding with increasing $f_0$, which leads to incomplete formation of shear bands. However, at larger strains, the volume fraction curve becomes monotonically increasing, and the effect of increasing $f_0$ is evident in the increased region occupied by the shear bands.}
\label{EvolutionSB}
\end{figure*}

\SK{The subsequent stress drop is no longer abrupt, unlike in the passive case at the same shear rate. A more gradual stress drop indicates a progressive failure with multiple stress relaxation events occurring simultaneously, resulting in more ductile-like behavior. The increase in yield stress occurs alongside a slight increase in the slope of the linear regime of the stress-strain curve, suggesting that adding activity raises the overall stress of the system.}
The steady-state stress at a fixed strain rate also rises. This is a counterintuitive observation, given that activity is often considered analogous to an effective temperature \cite{mandal2016active, PARRY2014183}. It is well-documented that increasing the temperature in a thermal passive system reduces both the peak stress and the steady-state stress. For a thermal passive system at a given strain rate, the yield stress ($\sigma_Y$) decreases with temperature \cite{PhysRevLett.95.195501, PhysRevB.87.020101} as follows:
\begin{equation}
\sigma_Y(T) = \sigma_Y^0 - cT^{2/3},    
\end{equation}
where $c$ is a constant, which will, in principle, depend on the strain rate $\dot\gamma$. A simple extension of this relation to effective temperature will naturally mean that with increasing activity, the effective temperature will increase and the yield stress will decrease. \JH{However, with increasing activity, the yield stress increases when the persistence time is sufficiently small. At larger persistence time, the concept of an effective temperature seems to be more appropriate, as dicussed later.} 

The system in the presence of activity shows a definite increase in transient steady-state stress with increasing $f_0$, as shown in Fig.~\ref{Rounded_Yielding}. \JH{To quantify this increase, we perform simulations on a smaller system ($N = 2000$) in two and three dimensions. For this system size, we find that a strain of the order of $\gamma = 10.0$ is required to achieve steady state. Importantly, we observe that both the stress and potential energy stabilise only in three dimensions, whereas in a two-dimensional system, the potential energy does not reach a steady state for a passive system and lower $f_0$ values under the same conditions (see Fig.~3 of supplementary material). In 3D, we observe that the steady-state stress increases with increasing $f_0$ as follows:}
\begin{equation}
  \sigma_{ss} (\dot{\gamma}, f_0) = \sigma_{ss}^{P}(\dot{\gamma}) + \alpha_{s1} f_0^{\beta_{s1}}, 
   \label{flowcurve_3d_N2k}
\end{equation}
where $\alpha_{s1}$ is the consistency parameter and $\beta_{s1}$ is the flow index, and both are strain-rate-dependent quantities. The first term $\sigma_{ss}^{P}$ represents the steady-state stress of the corresponding passive system (without activity). From the fitting, we extract $\beta_{s1} < 1$, confirming shear-thinning behavior in the active regime (represented by the solid lines). These observations are specific to systems with low persistence time. In the following sections, we explore how the system's behavior changes with varying $\tau_p$ values.

\vskip +0.1in
\noindent{\bf \large Activity-Induced Multiple Shear Bands:}
\JH{In amorphous solids under shear, shear band formation after yielding is a common feature. In ultrastable glasses, it is associated with a sharp stress drop in the stress-strain relation, as observed above for $f_0=0$ (see Fig.~\ref{Rounded_Yielding}). As we shall see now, the rounded yielding, observed with the inclusion of active or self-propelled particles (SSPs), is linked to the formation of multiple shear bands in both 2D and 3D.}
\SK{We have calculated $D^2_{min}$ of the particles to study the shear bands in the presence of SSPs at a strain rate of $\dot{\gamma} = 5\times 10^{-5}$ and $\tau_p=0.1$. Fig.~\ref{MultipleSB}(a) displays snapshots of shear bands at a fixed $\gamma = 0.25$ for 3D and $\gamma = 0.20$ for 2D with increasing $f_0 = 0.0, 1.0, 2.0, 4.0, 8.0$. A single shear band is observed in the passive case at $f_0 = 0.0$. As $f_0$ increases, the shear band grows and branches into multiple segments, which eventually develop into interconnected networks of shear bands. With further increasing $f_0$,} \JH{a more and more homogeneous flow pattern is observed.} 
\SK{Following the red region from left to right, we observe that as $f_0$ increases, the red region will increase and cover more space than the blue regions (more clearly seen in 2D). The red regions represent higher non-affine movements, with multiple plastic regions throughout the system as $f_0$ increases.} We show a similar increase in the shear-banded region with increasing $f_0$ for different strain rates ($\dot{\gamma} = 1\times10^{-4}, 1\times10^{-5}$) in Supplementary Fig.~5 and Fig.~6.

For better quantification, we calculate the volume/area fraction ($\phi$) occupied by particles exhibiting higher non-affine displacements, as shown in red (see Methods section for details). \JH{In Fig.~\ref{MultipleSB}(b) and (c), the volume fraction $\phi$ is shown as a function of the strain $\gamma$ for different choices of $f_0$ at $\tau_p = 0.1$}. In all the cases, the system begins at $\phi = 0$, indicating no shear band. \JH{For the passive system ($f_0 = 0$), there is a discontinuous jump around $\gamma \approx 0.12$, i.e.~around the location of the stress drop in the corresponding stress-strain relation (see Fig.~\ref{Rounded_Yielding}).} In contrast, active systems exhibit a continuous increase of $\phi$ with deformation. As $f_{0}$ increases, the strain at which $\phi$ first becomes non-zero shifts to larger values, signifying a delayed onset of shear‑band formation consistent with the increased yield strength observed in Fig.~\ref{Rounded_Yielding}. However, as strain progresses, the trend of increase in volume fraction reverses with $f_0$; the volume fraction for larger $f_0$ increases more steeply and saturates to a larger value because of the denser network structure formed at larger $f_0$, as visible to the right in Fig.~\ref{MultipleSB}(a). For the largest activity examined ($f_{0}=8.0$), $\phi$ remains zero to the greatest onset strain and then rises most steeply, reaching the largest value at high strains. We show this more clearly in the insets of Fig.~\ref{MultipleSB}(b) and (c), where the peak keeps shifting to larger strains and becomes increasingly broad, indicating a more gradual change under deformation with increasing active force.


\SK{Figure~\ref{EvolutionSB} compares the formation and evolution of shear bands in the passive system with the active system at the same strain rate $\dot{\gamma} = 5\times 10^{-5}$ in 3D and 2D.} The formation of the shear band is abrupt and system-spanning as it transitions from $\gamma = 0.12$ to $\gamma = 0.13$ in 3D and from $\gamma = 0.06$ to $\gamma = 0.07$ in 2D in the absence of activity. However, in the active case, Figs.~\ref{EvolutionSB}(b) and (d) show a gradual formation of multiple shear bands at $f_0 = 2.0$ from $\gamma = 0.11$ to $\gamma = 0.25$ in 3D, and at $f_0 = 4.0$ from $\gamma = 0.08$ to $\gamma = 0.20$ in 2D, at $\tau_p = 0.1$. \SK{Shear bands nucleate as localized deformation zones, forming isolated clusters that are initially sparse and fragmented, as shown in red. As illustrated, the red region grows with increasing strain at the expense of the blue region as deformation progresses. It is important to note that we do not see any clustering of active particles during shear band formation or proliferation. Rather, active particles are distributed homogeneously throughout the system. As the system reaches near the yield strain ($\gamma \approx 0.09$ in 2D  and $\gamma \approx 0.14$ in 3D), we observe multiple shear deformation zones that grow and interconnect forming a percolated network of shear bands spanning the system, nearly around $\gamma \approx 0.25$ in 3D and $\gamma \approx 0.20$ in 2D. The gradual formation of shear bands explains the continuous yielding observed in the stress-strain curve for the active cases (Fig.~\ref{Rounded_Yielding}). If further deformation is applied, the system continues to evolve, with an increased volume covered by the shear band, which can be quantified by calculating the volume fraction as shown in Figs.~\ref{EvolutionSB}(e) and (f). This parallels the effects of increased strain rates in the passive system, as the system requires more stress to activate multiple deformation pathways.}

\SK{The total volume fraction of plastic regions of the system also increases with increasing strain, as shown in Figs.~\ref{EvolutionSB}(e) and (f). At low strain, the volume fraction remains $0$ for all $f_0$ values. With increasing strain, we observe a first jump for the passive case, which yields the earliest, but for all other $f_0$ values, the volume fraction values remain smaller and closer to $0$. As strain further increases, there is a non-monotonic increase in the volume fraction with $f_0$ because for higher $f_0$, the yield point shifts to a larger value. Since shear bands interconnect continuously with applied deformation, we do not observe fully developed interconnected networks of shear bands for large $f_0$ at lower strain values. This happens at larger strains, leading to a systematic increase in volume fraction with increasing $f_0$.} This trend continues as the shear-banded region progressively grows with increasing strain. \SK{As shown in Fig.~\ref{EvolutionSB}, for a given strain greater than $0.4$ and $0.2$ in 3D and 2D, respectively, increasing $f_0$ has the same effect on the volume fraction of the region occupied by shear bands as being at a particular $f_0$ and increasing strain (Fig.~\ref{MultipleSB}), albeit for different reasons.} In the first case, the shear bands grow with strain, while in the second case, the network of shear bands becomes denser.

\SK{The formation of multiple shear bands occurs because high strain rates inhibit the propagation of a single system-spanning shear band, as the deformation occurs faster than the shear band propagation time, leading to independent shear bands forming at different sites \cite{PhysRevMaterials.4.025603}. When active particles are added to such a system, they enhance this effect, creating additional hindrance to shear band propagation and promoting the formation of multiple shear bands in response to external stress.}

\begin{figure*}[!htpb]
\includegraphics[width=0.99\textwidth]{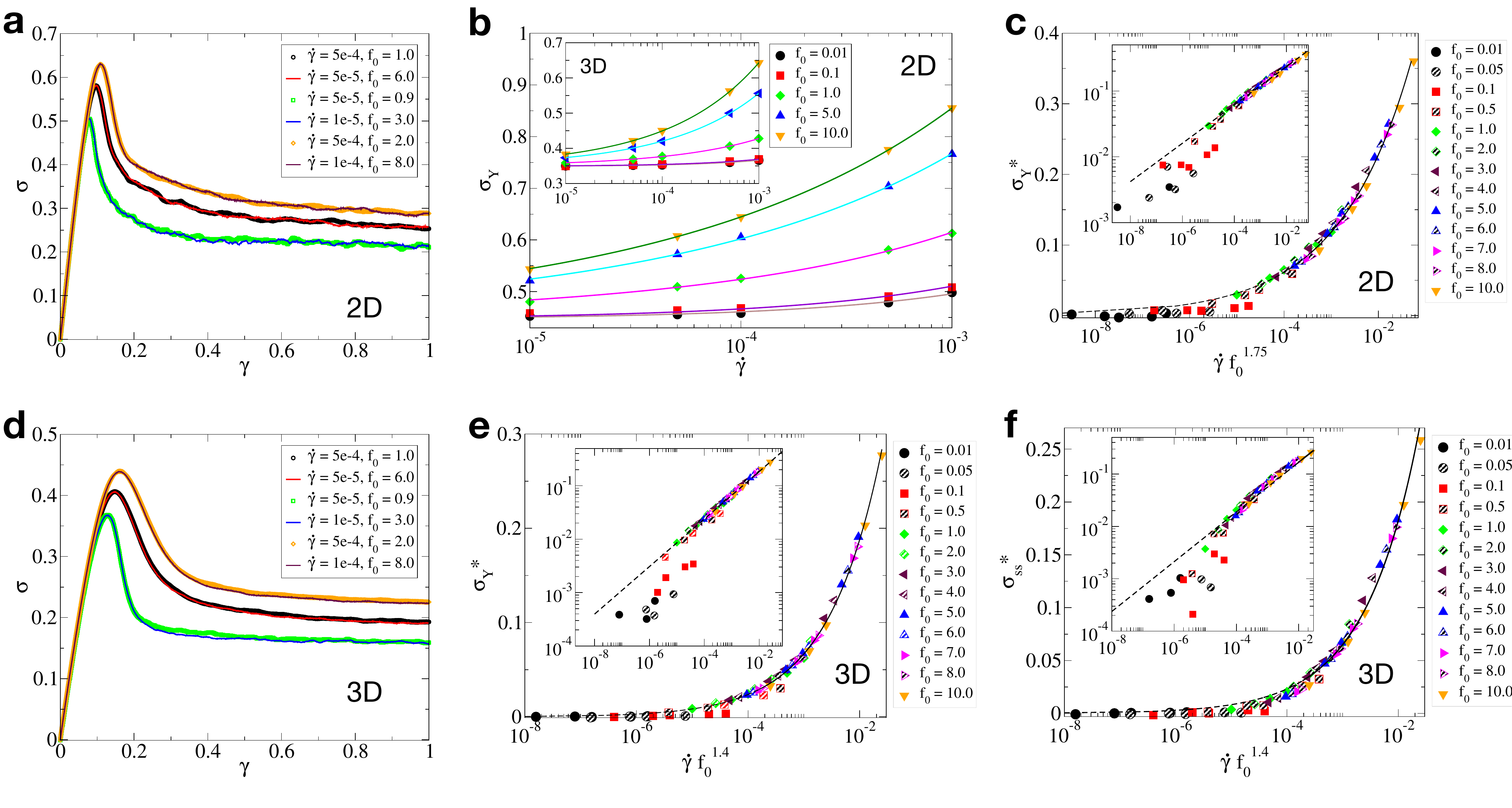}
\caption{{\bf Strainrate equivalence with active forces:} Panels (a) and (d) illustrate the equivalence between two cases in 2D and 3D: low strain rate with high $f_0$ and high strain rate with low $f_0$, demonstrating how different combinations of strain rates and active forces can produce a similar mechanical response. A system subjected to a faster strain rate $\dot{\gamma} = 5 \times 10^{-4}$ and lower active force $f_0=1.0$ exhibits identical mechanical response to that of a system with a slower strain rate $\dot{\gamma} = 5 \times 10^{-5}$ and higher active force $f_0=6.0$, regardless of dimensionality. Panel (b) plots the yield stress versus strain rate for several $f_{0}$ values; the data points (symbols) are fitted with Eq.~(\ref{Strainrate_f_rel2}) (solid curves). The critical $\dot{\gamma}$ required for a given $f_0$ to observe an increase in yielding decreases with an increase in $f_0$, while the critical $f_0$ needed decreases as $\dot{\gamma}$ increases (Figs.~\ref{Rounded_Yielding}(b), (e), and (h)), underscoring the compensatory relationship between these two control parameters. Panels (c), (e), and (f) show that the combination $\dot{\gamma}f_{0}^{m}$ collapses the yield stress (2D with $N=10000$ in (c) and 3D with $N=100000$ in (e)) and the steady-state stress (3D with $N=2000$ in (f)) onto a single master curve, implying that the driving mechanism of the system arises from a combination of strain rate and active forces. Additional values $f_{0}=0.05, 0.5, 2, 3, 4, 6, 7,$ and $8$ (beyond those used in (b)) are included to improve visualization. The collapsed data are fitted with Eq.~(\ref{Strainrate_f_combination}) (black dotted line). The insets show the data on log–log axes, revealing deviations from the fit at very small $\dot{\gamma}f_{0}^{m}$ values.}
\label{StrainRateF0_equiv}
\end{figure*}

\begin{figure*}
\includegraphics[width=0.99\textwidth]{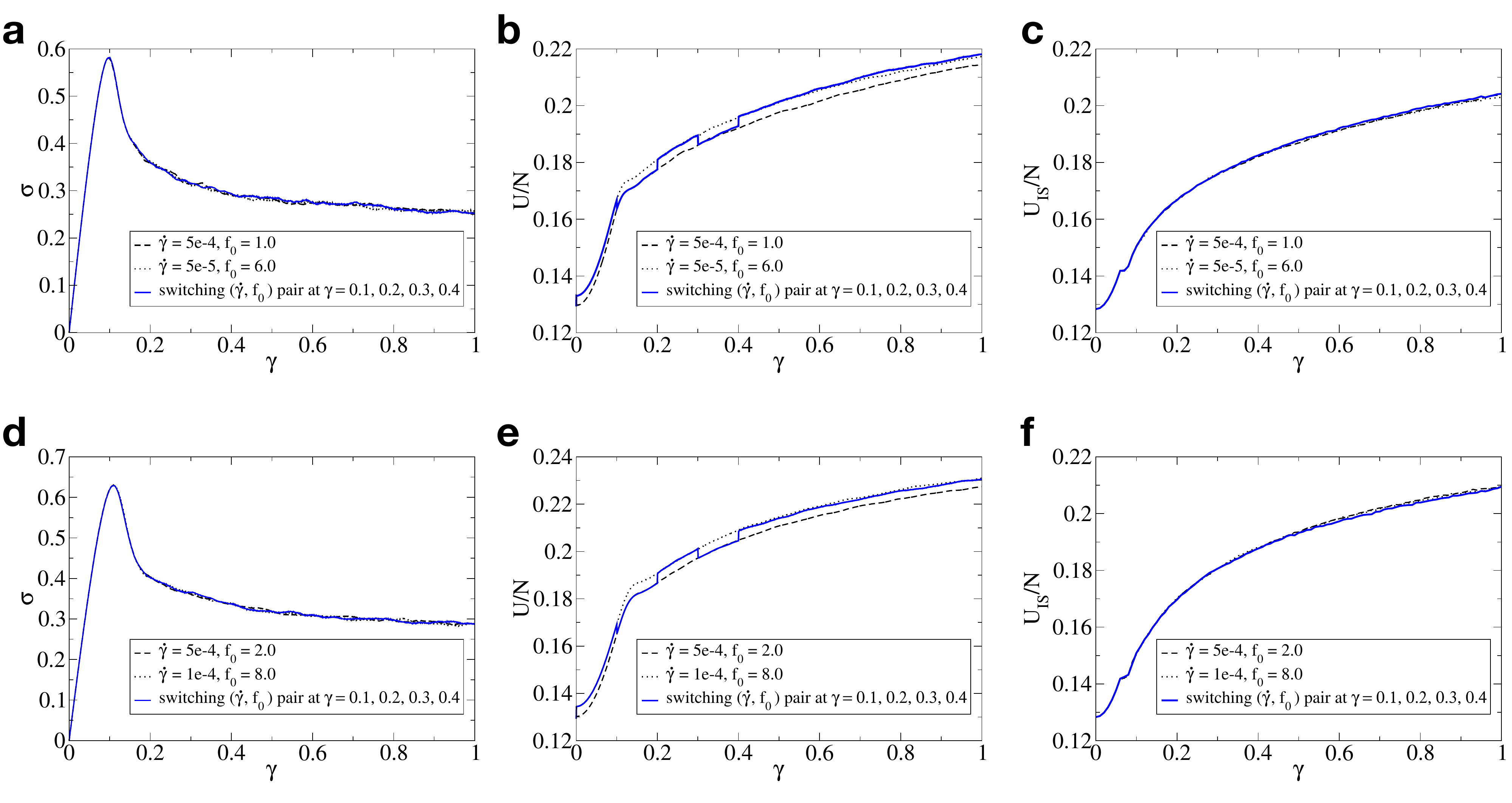}
\caption{{\bf Isomorphic behavior:} Panels (a) and (d) present stress-strain curves for 2D systems where pairs of $f_0$ and $\dot{\gamma}$ are switched at strains $\gamma = 0.1, 0.2, 0.3, 0.4$, demonstrating overlap with their originally equivalent pairs. Figures (b) and (e) illustrate the switching protocol: In (b), a jump in potential energy is observed at $\gamma = 0.10$ when switching from $\dot{\gamma} = 5 \times 10^{-5}$, $f_0 = 6.0$ to $\dot{\gamma} = 5 \times 10^{-4}$, $f_0 = 1.0$. Another jump occurs at \( \gamma = 0.20 \) when the system switches back from $\dot{\gamma} = 5 \times 10^{-4}$, $f_0 = 1.0$ to $\dot{\gamma} = 5 \times 10^{-5}$ , $f_0 = 6.0$, and so forth. Similar behavior is observed in (e) for different combinations of $f_0$ and $\dot{\gamma}$. The overlap of inherent state energies throughout the trajectory, as seen in (c) and (f), confirms the isomorph nature of the two curves. As a corollary, we also see that different combinations of $f_0$ and $\dot{\gamma}$ can be used to rejuvenate the glass to the same parent temperature (indicated by identical $U_{IS}$).}
\label{Isomorphism}
\end{figure*}

\begin{figure*}
\includegraphics[width=0.99\textwidth]{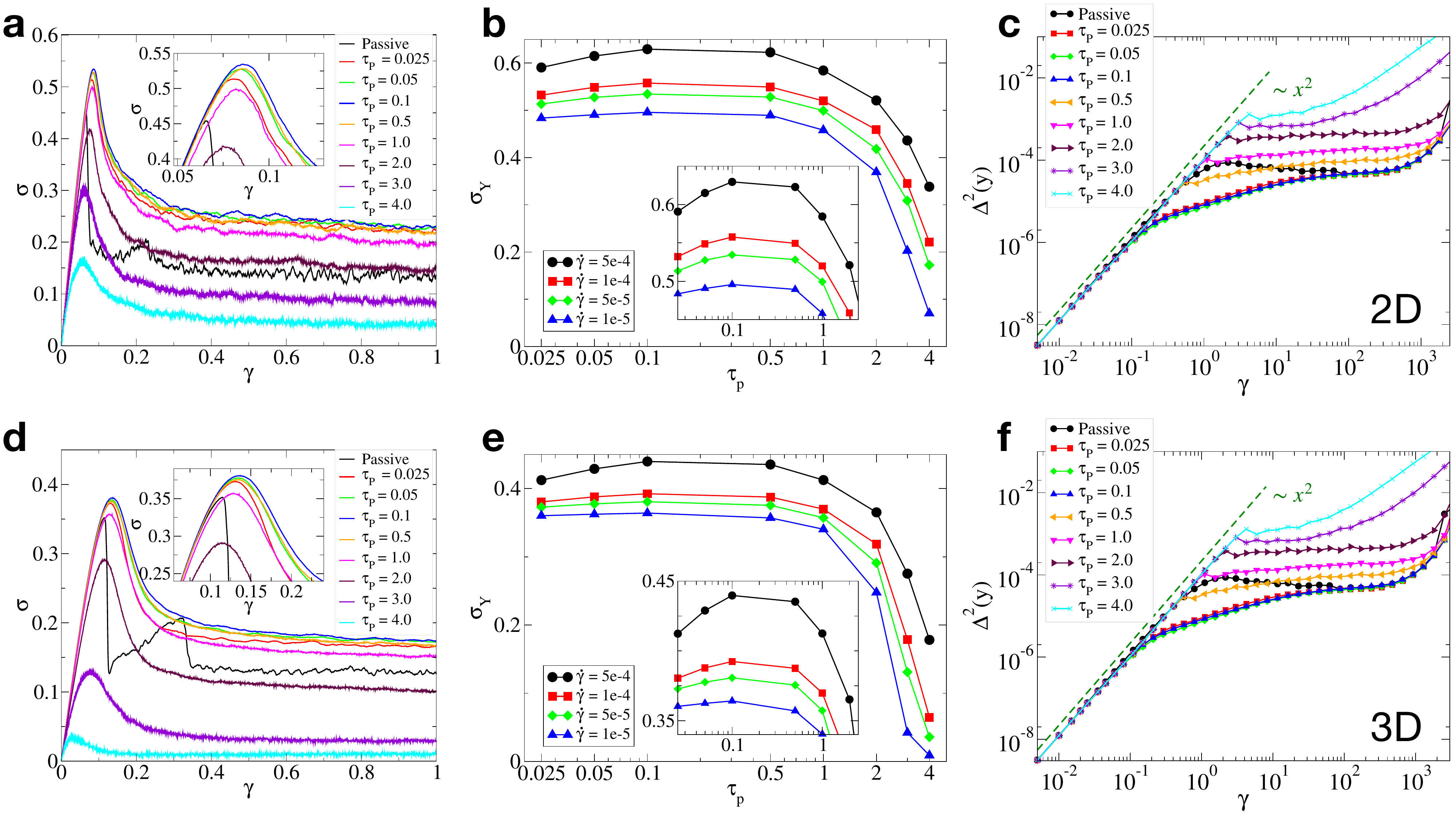}
\caption{{\bf Non-monotonic behavior:} As observed, for all strain rates, when $\tau_p$ is increased, the yield stress initially rises until $\tau_p = 0.1$, after which it decreases with further increases in $\tau_p$. The figures shown in (a),(b) and (d),(e) are shown for $f_0 = 2.0$ and strain rate $\dot{\gamma} = 5\times 10^{-5}$, where top panel is for 2D and bottom is for 3D, respectively. The non-monotonicity is evident in panels (b) and (e), where the yield stress is plotted as a function of $\tau_p$ for different strain rates, $\dot{\gamma} = 5\times 10^{-4}, 1\times 10^{-4}, 5\times 10^{-5}, 1\times 10^{-5}$. Since the value of $\tau_p$ at which the non-monotonicity occurs remains consistent across all cases—independent of strain rate, system size, or dimension, it is likely related to an intrinsic timescale of the system. The Mean-square displacement (MSD) of the system is shown in (c) and (f), and we observe that there is indeed a connection between the non-monotonic behavior of $\tau_p$ and the system's timescale, where the system transitions from ballistic motion into the cage regime. All curves follow ballistic motion until $\approx 0.1$, the system with $\tau_p \leq 0.1$ enters the cage earliest around time $0.1$, which continues increasing in a sub-diffusive manner. The value of $\tau_p$, for which this time is less than that of the passive (the curves which lie in between black and blue curves), shows an increase in mechanical stability, which is strain rate dependent. Beyond that, increasing $\tau_p$ leads to cage breaking, and particles diffuse without getting stuck in the cage, resulting in homogeneous plastic events taking place all over the system.}
\label{Nonmononicity}
\end{figure*}


\begin{figure*}
\includegraphics[width=0.99\textwidth]{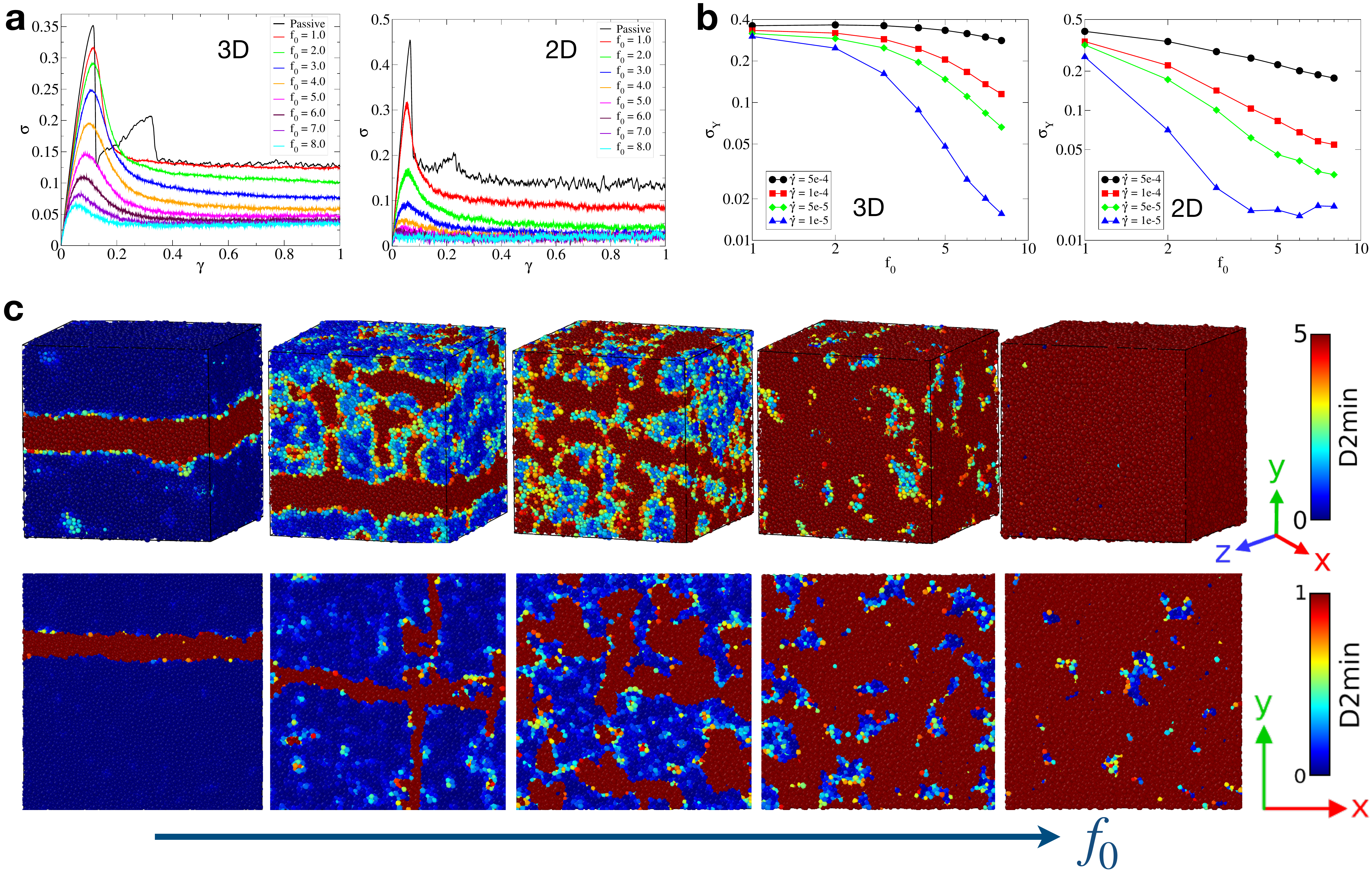}
\caption{{\bf Behavior of the system at higher persistence time:} The stress-strain curve shows a decrease in yield stress with an increase in $f_0$ for the systems studied in 3D and 2D at $\tau _p = 2.0$ and $\tau _p = 4.0$ as shown in (a). We see a monotonically decreasing curve if we plot the yield stress as shown in (b). We see complete fluidization as we increase $f_0$ at a strain rate of $\dot{\gamma} = 5\times 10^{-5}$ in 3D and 2D. In (c), the top panel shows the shear band at $\tau _p = 2.0$ in 3D, and the bottom row shows for $\tau _p = 4.0$ in 2D at $f_0 = 0.0, 1.0,2.0, 4.0, 8.0$. }
\label{HighertauData}
\end{figure*}

\vskip +0.1in
\noindent{\bf\large Compensatory Effects of Active Forces and Global Strain Rates:}
\JH{We have observed that at smaller persistence times the yield stress systematically increases with increasing active force $f_0$ at different strain rates. This hints at a possible relationship between strain rate and active force. At faster strain rates ($\dot{\gamma} = 1\times 10^{-4}$), multiple shear bands form even at low $f_0$, whereas slower strain rates ($\dot{\gamma} = 1\times 10^{-5}$) require higher $f_0$ to achieve the same effect (see Figs.~5 and 6 of the supplementary material). This compensatory relationship is quantitatively demonstrated by overlapping stress-strain curves for different combinations of $f_0$ and $\dot{\gamma}$ with equivalent yielding and steady-state responses. For instance, at $\dot{\gamma} = 5 \times 10^{-4}$ with $f_0 = 1.0$, similar values of yield stress and yield strain ($0.58$ and $0.097$, respectively) are seen as those observed at $\dot{\gamma} = 5 \times 10^{-5}$ with $f_0 = 6.0$ ($0.58$ and $0.098$, respectively). This behavior is illustrated in Figs.~\ref{StrainRateF0_equiv}(a) and (d), where the stress-strain curves in both 2D and 3D demonstrate that shear deformation and active forces exhibit compensatory effects.}

\JH{To quantify these effects, we first assume that the strain-rate dependence of the yield stress in the passive case, $\sigma_Y^P(\dot\gamma)$, can be expressed as 
\begin{equation}
\sigma_Y^P(\dot\gamma) = \sigma_{AQS}^P + \alpha_0 \dot{\gamma}^{\beta_0} 
\label{eq_sigmayp}
\end{equation}
where $\sigma_{AQS}^P$ is the yield stress for the passive system in the AQS limit, i.e.~at zero temperature in the quasistatic limit. The consistency index $\alpha_0$ and the flow index $\beta$ in Eq.~(\ref{eq_sigmayp}) are assumed to be independent of the strain rate. By inserting Eq.~(\ref{eq_sigmayp}) into Eq.~(\ref{HBstress}), we obtain the following equation:}
%
\begin{eqnarray}
\sigma_Y (\dot\gamma, f_0) = \sigma_{AQS}^{P} +  \alpha_0 \dot{\gamma}^{\beta_0} + {\alpha_1} f_0^ {\beta_1}.
\label{Strainrate_f_rel1}
\end{eqnarray}
\JH{Here, both $\alpha_1$ and $\beta_1$ depend on strain rate. For $\alpha_1$, we assume a power-law dependence on $\dot\gamma$, $\alpha_1 = \alpha \dot{\gamma}^\beta$, where $\alpha$ and $\beta$ are constants that in particular are independent of strain rate and the active force $f_0$. Using this relation for $\alpha_1$, we can write}
%
\begin{eqnarray}
\sigma_Y(\dot\gamma,f_0) = \sigma_{AQS}^P +  \alpha_0 \dot{\gamma}^{\beta_0} + \alpha \dot{\gamma}^{\beta} f_0^{\beta_1} \, .
\label{Strainrate_f_rel2}
\end{eqnarray}
\JH{In Fig.~\ref{StrainRateF0_equiv}(b), we show the yield stress as a function of strain rate for different values of $f_0$. This plot indicates a shear-thinning behavior, similar to the case of $\sigma_Y$ as a function of $f_0$ at different strain rates (Fig.~\ref{Rounded_Yielding}). The solid lines in Fig.~\ref{StrainRateF0_equiv}(b) are fits to Eq.~(\ref{Strainrate_f_rel2}). We find ${\beta} < 1$ for all cases in both 2D and 3D. The plot shows that with increasing $f_0$, the yield stress increases rapidly, showing similar behavior to previous observations, albeit with an interesting difference; now, the yield stress exhibits a more pronounced increase at small $f_0$ for higher $\dot{\gamma}$, or at small $\dot{\gamma}$ with larger $f_0$. This interchangeable behavior enables data collapse through a simple rescaling, as shown in Figs.~\ref{StrainRateF0_equiv}(c) and (e), where the yield stress data collapses onto a single line when the $x$-axis is rescaled to $\dot{\gamma} f_0^m$. The resulting curve can be described using Eq.~(\ref{Strainrate_f_rel2}) with setting $\beta_1 = m \beta$ in the third term:}
%
\begin{eqnarray}
\sigma_Y(\dot\gamma, f_0) =  \sigma_{AQS}^P +  \alpha_0 \dot{\gamma}^{\beta_0} + {\alpha} ({\dot{\gamma}f_0^m}) ^ {\beta},
\label{Strainrate_f_combination}
\end{eqnarray}
The term ${\alpha} ({\dot{\gamma}f_0^m})^{\beta}$ captures the non-linear coupling between activity and shear rate, where ${\alpha}$ and ${\beta}$ are constants with $m {\beta} \approx {\beta_1}$. The value of the fitting parameter $m$ is $1.75$ in 2D and $1.4$ in 3D. The exponent ${\beta} \approx 0.29$ in 2D and ${\beta} \approx 0.45$ in 3D, making ${\beta_1} \approx 0.5$ in 2D and ${\beta_1} \approx 0.63$ in 3D. These exponents indicate shear-thinning behavior under both increasing active forces and shear rates. We further define the scaled yield stress as $\sigma_Y^* = \sigma_Y(\dot\gamma, f_0) - \sigma_{AQS}^P - \alpha_0 \dot{\gamma}^{\beta_0}$, which is plotted in Fig.~\ref{StrainRateF0_equiv}(c) and (e) for 2D and 3D, respectively.

We observe a similar collapse for steady-state stress in 3D for a system size of $N=2000$, in a similar way by putting $\sigma_{ss}^P (\dot{\gamma}) = \sigma_{ss}^P +  \alpha_{s0} \dot{\gamma}^{\beta_{s0}}$ and expanding $\alpha_{s1} = \alpha_{s} \dot{\gamma}^{\beta_s}$.
\begin{eqnarray}
\sigma_{ss}(\dot\gamma, f_0) =  \sigma_{ss}^P +  \alpha_{s0} \dot{\gamma}^{\beta_{s0}} + {\alpha_{s1}}f_0^{\beta_{s1}}\\
\sigma_{ss}(\dot\gamma, f_0) =  \sigma_{ss}^P +  \alpha_{s0} \dot{\gamma}^{\beta_{s0}} + {\alpha_s} ({\dot{\gamma}f_0^{m_s}}) ^ {\beta_s},
\label{Strainrate_f_combination_steadystate}
\end{eqnarray}

We scale the steady-state stress as follows: $\sigma_{ss}^*$ is defined as $\sigma_{ss}^* =\sigma_{ss}(\dot\gamma, f_0) - \sigma_{ss}^P - \alpha_{s0} \dot{\gamma}^{\beta_{s0}}$, with $m_s = 1.4$, and the dotted line shown in Fig.~\ref{StrainRateF0_equiv}(f) corresponds to $\sigma_{ss}^* = {\alpha_s} ({\dot{\gamma}f_0^{1.4}})^ {\beta_s}$, with $\beta_s < 0.5$ ($\beta_s \approx 0.45$). The insets shown in (c), (e) and (f) show the same scaled data in a log-log plot for the same $f_0$ and $\dot{\gamma}$ values.

\SK{This scaling and data collapse reveals a fundamental description of the mechanical response of the system where the product $\dot{\gamma} f_0^m$ serves as a scaling parameter. This implies that the system's mechanical response to deformation is inherently linked to the activity. The collapse for different values of $\tau_p$ is shown in Fig.~7 of the supplementary material.} 

\JH{Although $f_0$ and $\dot{\gamma}$ have conceptually very different behaviors and can be controlled independently in the simulation, Eqs.~(\ref{Strainrate_f_combination}) and (\ref{Strainrate_f_combination_steadystate}) show that there can be occasions where various combinations of $\dot{\gamma}$ and $f_0$ will lead to the same yielding response, suggesting a possible isomorphic behavior. These similarities indicate an underlying fundamental mechanism that is at play for identical mechanical responses. In Figs.~\ref{Isomorphism}(a) and (d), we have shown the stress-strain curve for a combination of $\dot\gamma$ and $f_0$ such that $\alpha_0 \dot{\gamma}^{\beta_0} + {\alpha} ({\dot{\gamma}f_0^m}) ^ {\beta}$ is nearly constant and we have indeed found that the resulting stress response is identical to each other. However, when we plot the potential energies of two overlapping stress-strain curves, they were found to be different. In Figs.~\ref{Isomorphism}(b) and (e), we show the potential energy curves corresponding to the equivalent stress-strain responses shown in panels (a) and (d) as indicated by the dotted lines.}

\SK{To investigate this further, we devised a protocol: the shear was initiated with ($\dot\gamma, f_0$) = ($5\times 10^{-5},6.0$), and the first switching to ($5\times 10^{-4},1.0$)  was performed at a strain $\gamma = 0.1$. Then again, in the second switching, the system returned its original dynamical condition of  $\dot{\gamma} = 5\times 10^{-5}$, $f_0 = 6.0$ at a strain of $\gamma = 0.2$.  This switching process was repeated.  The third switch from ($\dot{\gamma} = 5\times 10^{-5}$, $f_0 = 6.0$) to ($\dot{\gamma} = 5\times 10^{-4}$, $f_0 = 1.0$) at $\gamma = 0.3$ and the final switch from ($\dot{\gamma} = 5\times 10^{-4}$, $f_0 = 1.0$) back to ($\dot{\gamma} = 5\times 10^{-5}$, $f_0 = 6.0$) at $\gamma = 0.4$. Notably, at each switch, a jump is observed in the potential energy curve, depicted by the blue color curve in Figs.~\ref{Isomorphism}(b) and (e). The dotted lines indicate the set of $\dot{\gamma}$ and $f_0$ values between which the switching was performed.}

\SK{Note that the jumps observed in Figs.~\ref{Isomorphism}(b) and (e) are absent in the stress-strain curves (a), (d) as well as in the inherent state energies shown in panels (c) and (f).} To verify whether this behavior persists in the steady state, we examine a smaller system size ($N=2000$) and show Fig.~9 of the supplementary material that the potential energy remains different between isomorphic pairs, while the stress and inherent state energy remain identical. If one plots these isomorphic stress-strain curves for different ensembles, the yielding response for the same ensemble is the same, and they fluctuate around the same steady-state stress, as shown in Fig.~8 of the supplementary material. Additionally, the difference between the potential energy and the inherent‑state energy shows a jump at the switching points and otherwise remains constant within the same state, as illustrated in Fig.~10 of the supplementary material.
\SK{This indicates that the macroscopic response to shear and activity remains the same throughout shear, regardless of these jumps confirming the isomorphic nature of the two overlapping curves. This consistency arises because the respective pairs of strain rate and active forces drive the system toward the same inherent state, resulting in similar structural responses. This is a significant result, as it demonstrates that the compensatory effect of global shear and activity can be effectively utilized to rejuvenate ultrastable glass to the same parent temperature. For instance, in scenarios where it is difficult to control external shear (e.g., confined geometries or in certain experimental setups), one can tune internal activity to achieve the same rejuvenation and mechanical response. Conversely, if controlling activity is challenging (e.g., in systems with limited active particle concentration), adjusting the external shear rate can produce equivalent effects.}


\vskip +0.1in
\noindent{\bf\large Non-Monotonic Yielding with Persistence Time:}
\SK{As persistence time, which influences how an active system explores the configurational space, is another important aspect of any active system, we now focus on how the effect of persistence time, $\tau_p$, on yielding differs from that of active forces, $f_0$. It plays an important role in the increase or decrease of yield stress, providing a window to optimize activity for desired behavior. As shown in Fig.~\ref{Nonmononicity}, when $\tau_p$ is increased, the yield stress initially increases until $\tau_p \sim 0.1$ and then decreases with further increases across all strain rates studied. Panels (a) and (d) show stress-strain curves for 2D systems with $N = 10000$ and 3D systems with $N = 100000$ at a strain rate of $\dot{\gamma} = 5\times 10^{-5}$, respectively. The non-monotonic behavior is evident in panels (b) and (e), where the yield stress is plotted as a function of $\tau_p$ for different strain rates. The value of $\tau_p$ at which this non-monotonicity occurs remains nearly constant at around $0.1$ across all cases, independent of strain rate, system size, or dimensionality (see also Fig.~11 of the supplementary material). This consistency across different conditions suggests that this characteristic $\tau_p$ value might be related to an intrinsic timescale of the system.} 

\JH{To gain a better understanding on how the time scale $\tau_p$ affects the dynamics, we now consider the mean square displacement (MSD, $\Delta^2$). It is shown in Figs.~\ref{Nonmononicity}(c) and (f) for different values of $\tau_p$. The figures indicate an interesting behavior of the MSD with respect to the length of ballistic regime and the height of the plateau in the caging regime. As $\tau_p$ increases from $0.1$ to $1.0$ in 3D and $0.1$ to $2.0$ in 2D, the system enters the caging regime earlier than the passive system, and this behavior is strain-rate-dependent. As shown in Fig.~12 of the supplementary material for 3D, at faster strain rates ($\dot{\gamma} = 5\times 10^{-4}$), the yield stress for $\tau_p \leq 2.0$ exceeds that of the passive system. However, at slower strain rates ($\dot{\gamma} = 1\times 10^{-5}$), this holds true only for $\tau_p \leq 0.5$. The maximum yield stress is consistently observed at $\tau_p = 0.1$, characterized by a shorter ballistic regime. Similar behavior is observed in 2D.}

\SK{As we increase persistence time (till $\tau_p = 1.0$), active particles maintain a slope of $2$ longer than passive particles around $t_{b2}$ but enter the cage quickly (slope $\approx 0$), which effectively increases the caged time of particles compared to the passive system, resulting in increased mechanical strength for these $\tau_p$ values. However, with further increasing $\tau_p$, the duration of the ballistic regime increases, and at sufficiently high $\tau_p$, the system transitions directly from ballistic to diffusive motion, escaping the cage. Increasing $f_0$ at a $\tau_p = 0.1$ only affects the diffusive regime, as shown in Fig.~13 of the supplementary material. For $\tau_p = 2.0$, the system's caging time reduces, and transitions from ballistic to diffusive become smoother. Two key observations emerge: first, the initial deviation from ballistic motion occurs around $t_{b1} \approx 0.1$, explaining the non-monotonicity; second, increased mechanical strength is observed until $t_{b2}$, where the active system enters caging earlier than the passive system. With further increasing $\tau_p$, the persistence time becomes large enough for active particles to break their cages and diffuse. Notably, in the passive system, the transition to the diffusive regime occurs discontinuously, coinciding with failure and shear band formation. However, with activity present, this transition becomes progressively smoother with increased diffusivity.}

\SK{The non-monotonic behavior in persistence time raises the question of how the mechanical response of the system changes with increasing $f_0$ as the persistence time continues to increase. In the Fig.~2 of the supplementary material, the yield stress as a function of $f_0$ at different $\dot{\gamma}$ for $\tau_p$ values $0.05, 0.1, 0.5, 1.0$ in 2D is presented. As $\tau_p$ increases, the behavior transitions from a monotonically increasing yield stress to a non-monotonic response, which becomes evident at $\tau_p = 1.0$. With further increase in the persistence time, the behavior shifts again to a monotonically decreasing yield stress with increasing $f_0$ values which is evident from the stress-strain curve in Fig.~\ref{HighertauData}(a) and yield stress in (b). A marked difference is observed in both 3D and 2D when compared to the systems with low persistence time (Fig.~\ref{Rounded_Yielding}). For higher $\tau_p$ as shown in Figs.~\ref{HighertauData}(a) and (b), for $\tau _p =2.0$ in 3D and $\tau _p = 4.0$ in 2D, increasing $f_0$ leads to a progressive reduction in yield stress, ultimately resulting in complete fluidization. Higher persistence times lead to cage breaking by active particles, as they have sufficient time to escape their local cages. This results in a more uniform stress distribution throughout the system, reducing both yield stress and steady-state stress as the material transitions to a more fluid-like state. In Fig.~\ref{HighertauData}(c), the snapshots illustrate the evolution of the shear band for $\tau_p = 2.0$ in 3D and $\tau_p = 4.0$ in 2D with increasing $f_0$ values $0.0, 1.0, 2.0, 4.0, 8.0$. At low $f_0$ values $0.0, 1.0$, system-spanning shear bands are observed because the forces are insufficient for the particles to break their local cages for the given persistence time. With increasing active forces ($f_0 \geq 2.0$), particles transition to a diffusive regime without getting stuck in their local cages, allowing for more global rearrangements. This results in more homogeneous plastic drops, facilitating easier flow and lower yield stresses, which ultimately leads to fluidization.}

\vskip +0.1in
\noindent{\bf\large Dependence of Shear Band Length Scale on Active Forces:}
\begin{figure}[ht]
\includegraphics[width=0.48\textwidth]{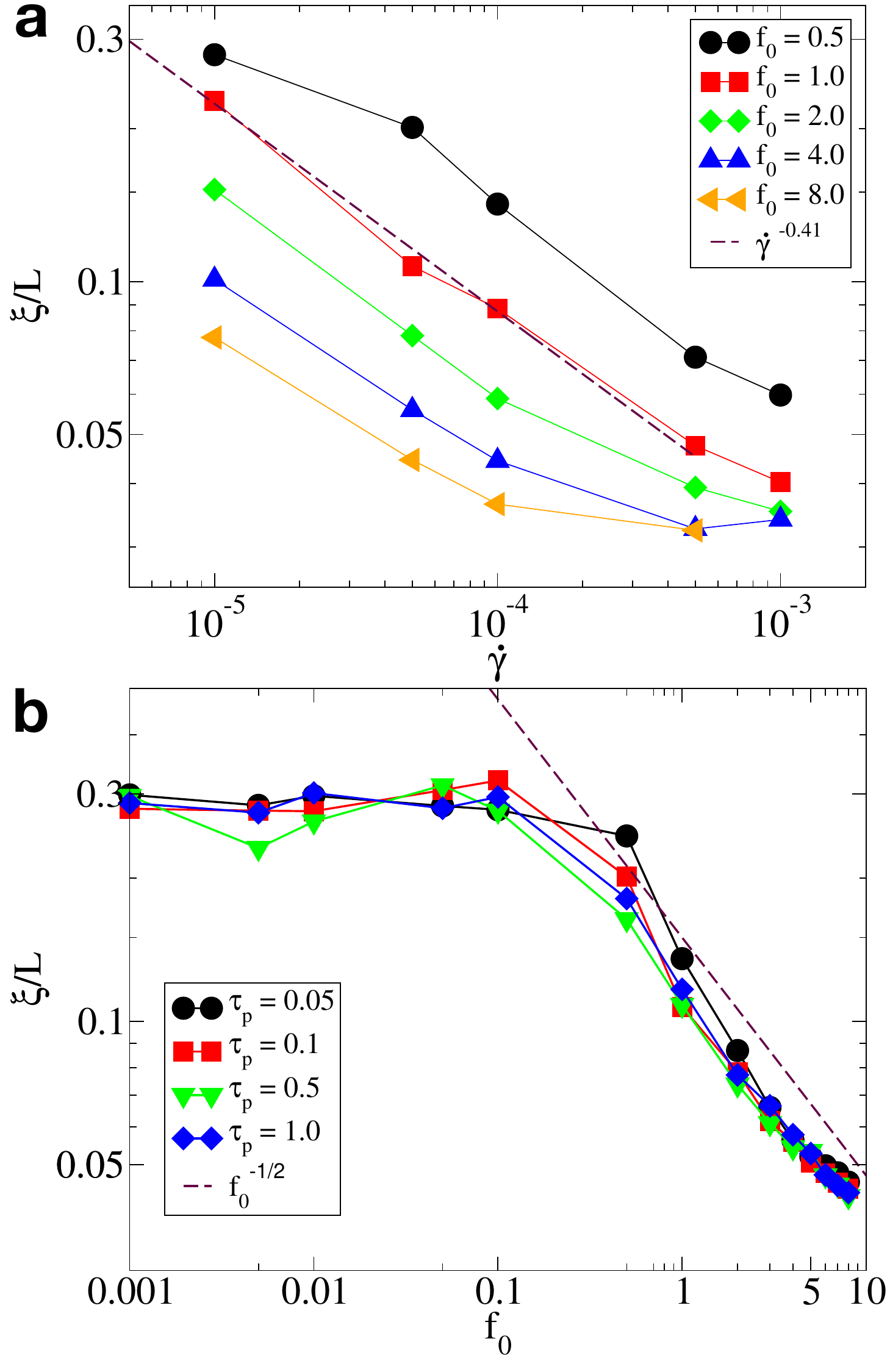}
\caption{{\bf Separation between shear bands:} \JH{(a) $\xi/L$ as a function of strain rate for different values of $f_0$. The decrease of $\xi(\dot{\gamma})/L$ follows Eq.~(\ref{eq_xi}) with $\alpha = 0.41$, obtained by fitting the data for $\tau_p = 0.1$ (dashed line). The distance between shear bands, $\xi/L$, is obtained at a strain $\gamma_A$ just after yielding for a system with $L=100$. (b) $\xi/L$ as a function of $f_0$ for different values of  $\tau_p$ (0.05, 0.1, 0.5, and 1.0) at the strain rate $\dot{\gamma} = 5\times 10^{-5}$. With increasing $f_0$, $\xi/L$ decreases in an $f_0^{-1/2}$ manner, implying an increase in the number of multiple shear bands, which leads to decreased separation between them. At lower values of $f_0$, it saturates to a value ($\approx 0.3$) due to the absence of multiple shear bands.}}
\label{Lengthscale}
\end{figure}
\JH{The observation of multiple shear bands with increasing $f_0$ suggests the emergence of an underlying length scale ($\xi$) related to inter shear band distance due to active forces. In Ref.~\cite{PhysRevMaterials.4.025603}, an interesting relationship between such a lengthscale and the shear rate has been proposed for passive systems. Here, we show that in the presence of active forces a similar relationship holds. To quantify this relationship, we first create a binary image of the shear-banded system following the same procedure as above for the calculations of the volume fraction of shear-banded regions. As in Refs.~\cite{PhysRevMaterials.4.025603, PhysRevLett.106.125702}, we draw horizontal and vertical lines to measure chord lengths between consecutive shear bands (see supplementary material for details).} \SK{After averaging the first moment of distance data across all ensembles, the evolution of $\xi/L$ with strain rate for both passive and active cases is shown in Fig.~\ref{Lengthscale}(a). We observe that the spacing between shear bands decreases algebraically with strain rate for the considered values of $f_0$. The data corresponding to $\tau_p = 0.1$ was fitted using
\begin{eqnarray}
\xi \sim \dot\gamma^{-\alpha}
\label{eq_xi}
\end{eqnarray}
with $\alpha \simeq 0.41$ as represented by a dashed line in panel (a). The exponent of the power law seems to be independent of $f_0$, although for larger $f_0$, the number of shear bands becomes too large to compute the spacing between the shear bands very reliably, and the data deviates from the expected power-law behavior.}

\SK{To quantify how active forces promote the formation of multiple shear bands under global shear, we plot $\xi/L$ as a function of $f_0$ at $\dot\gamma = 5\times10^{-5}$ for different persistence times $\tau_p = 0.05, 0.5, 0.1, 1.0$ in Fig.~\ref{Lengthscale}(b). The length scale is found to decrease with increasing $f_0$. For further quantification, we define a strain value $\gamma_A$ just after yielding, where $\Delta \sigma = \sigma_Y - \sigma_{ss}$ decays to $75\%$ of its initial value. For higher strain rates ($\dot\gamma = 5\times 10^{-4}, 10^{-3}$) and larger active forces ($f_0 = 8.0$), the stress continues to decrease without reaching steady-state up to $100\%$ strain deformation within the studied simulation window. For these cases, we use $\gamma_A = 0.26$ as chosen for $f_0 = 4.0$. At these parameters, the separation also becomes so small for the given system size that they deviate from the power law behavior. We observe that with increasing $f_0$, the spacing decreases for $f_0 \geq 0.5$ as 
\begin{equation}
\xi/L \sim f_0^{-1/2}.   
\end{equation}
At lower values ($f_0 \leq 0.1$), it saturates to a value ($\approx 0.3$) as there will be only a single shear band in the system at the yielding point. For higher $\tau_p$, the data is not shown as there are no defined shear bands as can be seen in Fig.~\ref{HighertauData}.}

\SK{It is well-known that the passive case exhibits a single system-spanning shear band in the ($\dot{\gamma} \rightarrow 0$) AQS limit. In the presence of finite shear rate and activity, the system shows a single shear band due to finite-size effects. Increasing the system size at low activity or shear rate will result in multiple shear bands separated by very large distances (larger than the system sizes studied here). What we want to highlight in this study is that doping the system with a small amount of activity at a slower shear rate decreases the separation between shear bands, similar to what one may obtain by increasing the strain rate to a very large value in the passive case. The separation length $\xi/L$ provides insight into the system's ductility. At very large separation length the system can be interpreted as being brittle, while a decrease in separation length indicates increasing ductility. This supports the conclusion that activity enhances the ductility of the system.}
 
\vskip +0.1in
\noindent{\bf\large Discussion:}
\SK{This study advances and adds another dimension to our understanding of a fundamentally challenging problem of heterogeneous, shear-band-mediated failure in ultrastable glasses by investigating yielding behavior and shear band formation in the presence of active forces. We demonstrate that incorporating self-propelled particles (SPPs) at finite temperatures and strain rates can enhance the mechanical stability of ultrastable glasses under homogeneous shear conditions. This enhancement is quantitatively characterized by a significant increase in yield stress and yield strain of the system. We also report multiple shear band formations with an increase in the magnitude of active forces at a fixed shear rate, and their gradual development accounts for the continuous yielding behavior observed in the stress-strain curve. The observation of this progressive failure mechanism post-yield point has significant implications for future research as well as industrial applications.}


\SK{Furthermore, our analysis shows that strain rate and active forces exhibit a complementary relationship, implying that a faster strain rate with low $f_0$ is equivalent to a slower strain rate with high $f_0$. This compensation arises from the isomorphic nature of a pair of combinations of strain rate and active force that rejuvenate the ultrastable glass to the same inherent state. In situations where external shear is challenging to control, internal activity can be adjusted, and when tuning activity puts constraints, modifying the external shear rate can yield the same effects. In biological scenarios, it may be more feasible for a system to regulate its internal activity rather than control external forces acting on it. The phenomenon of glass yielding under external stress can thus be mimicked in dense active glassy systems, offering insights into yielding in biological matter. This demonstrates the versatility of these mechanisms in tailoring the structural and dynamic properties of ultrastable glasses. Thus, the complementary relationship between strain rate and active forces suggests new directions for future research, where one can utilize this concept to explore biological systems under mechanical load.} 

\SK{The interplay between persistence time and the system's intrinsic timescale emerges as another critical factor, where particles move freely before experiencing caging effects. Shorter persistence times lead to an earlier onset of the caged regime, which results in a longer caged time and enhanced mechanical stability, as evident from pronounced stress peaks. Although this enhancement of mechanical stability does not lead to stronger strain localization, they initiate multiple shear localization zones, forming multiple shear bands as strain increases. The distance between shear bands decreases with increasing active forces, indicating that the system becomes more ductile by introducing active dopants at lower persistence times.  This has practical implications for experimental research on ultrastable glass and industries manufacturing ultrastable glasses, where brittleness has been a well-known limitation. Using activity as a dopant provides a potential solution to this challenge. At higher persistence times, on the other hand, the particles escape their cages, and as the active forces increase, this cage-breaking event is enhanced, leading the system to liquefy as $f_0$ increases at a fixed $\tau_p$ or $\tau_p$ increases. This behavior is evident in both stress-strain curves and shear band structures with increasing active forces. The two $\tau_p$ regimes thus exhibit opposing behaviors,  highlighting that tuning the system to the appropriate $\tau_p$ is critical for either enhancing the system’s mechanical stability or inducing fluidization under external forces.} 

\SK{Finally, this work contributes to the understanding of shear band formation in the presence of active forces. However, the microscopic mechanisms of how stress builds up and activates multiple deformation pathways remain an area for further investigation. Additionally, the roles of concentration and temperature in this context have not been addressed in this study. Tuning these parameters will significantly impact the yielding and formation of shear bands, making them important for future research.}

\section*{Methods}\label{sec:methods}
\subsection{MODEL \& SIMULATION DETAILS \protect\\}

\noindent{\bf Model:} 

Ninarello {\it et al.}~\cite{PhysRevX.7.021039, 10.1063/5.0086626} proposed a model system to study glasses that accelerates thermalization and prevents crystallization. In this model, polydisperse soft spheres interact through pairwise additive interactions, and the particle diameters exhibit non-additivity. Each particle's diameter in this model conforms to a probability distribution $P(\sigma)=A \sigma ^{-3}$ with $A=2/(\sigma_{min}^{-2}-\sigma_{max}^{-2})$. Following Refs.~\cite{PhysRevX.7.021039, 10.1063/5.0086626}, we chose the minimum and maximum diameters as $\sigma_{min}=0.725 \overline \sigma$ and $\sigma_{max}\approx 1.611 \overline \sigma$. The unit of length is set by mean particle diameter $\overline\sigma =1.0$. The interaction potential for the particles is defined as:
\begin{eqnarray}
V(r)=
\begin{cases}
V_0 (r^{-12} + C_0 + C_2 r^2 + C_4 r^4) &  ,r < r_c \\
0   & ,r \geq r_c
\end{cases}
\label{Potential}
\end{eqnarray}
where $r=r_{ij}/\sigma_{ij}$, the cut-off distance is $r_c = 1.25$, and $V_0 =1.0$ sets the energy unit in above equation. The other parameters in the potential $C_0$, $C_2$, and $C_4$ ensure the function's smoothness at the cutoff, with values $C_0 = -28/r_c^{12}$, $C_2 = 48/r_c^{14}$, and $C_4=-21/r_c^{16}$.

To investigate the impact of active forces on the shear band, we introduced an active force $\vec{f_0}$ to a fraction of the particles chosen randomly in the system. The active force modifies the equation of motion as follows:
\begin{eqnarray}
\boldsymbol{\dot{r}_i} &=& \frac{\boldsymbol{p}_i}{m}\\
\boldsymbol{\dot{p}}_i &=& -\frac{\partial \boldsymbol{V}}{\partial \boldsymbol{r}_i}+n_i \boldsymbol{f}_i,
\label{EOM}
\end{eqnarray}
with $n_i = 1$ for active particles and $0$ otherwise. The force $\vec f_0$ is defined as
\begin{eqnarray}
  \vec{f_0}=f_0(k_x \hat{x} +k_y \hat{y} + k_z \hat{z} ), 
  \label{Active}
\end{eqnarray}
where $k_x$, $k_y$, and $k_z$ control the direction in which any active particle moves. The value of $k_x$, $k_y$, and $k_z$ is chosen from $\pm 1$, which applies the force along eight diagonal directions, similar to the 8-state clock model. The sum of $k$ in all directions is maintained at zero to ensure momentum conservation. These active particles are self-propelled in a direction for a fixed amount of time, known as persistence time $\tau_p$. After every $\tau_p$, the direction of the active particles is randomized to mimic the run-and-tumble motion observed for various biological and synthetic active particles.

\noindent{\bf Sample Preparation:}
The system was equilibrated at a constant temperature of $T=0.06$ for the 3D model and $T=0.026$ for the 2D model. Equilibration was achieved for a system size of $N=2000, 10000$ in 2D and $N=2000, 10000, 100000$ in 3D using the hybrid MD-SMC method. In this method, a swap attempt was made every 25 MD steps with $N$ trial SMC moves, ensuring that the probability of swap attempts remains independent of system size. Simulations were performed under periodic boundary conditions with a time step of $dt=0.005$ and a number density of $\rho=1.0$. This density corresponds to a simulation box length of $L = 100$ in 2D for $N=10000$ and approximately $L \approx 46.42$ in 3D for $N=100000$. Unless otherwise specified, all 3D results are reported for systems with $N = 100000$ particles (averaged over eight ensembles) and $N = 2000$ particles (averaged over 32 ensembles), while 2D results are reported for $N = 10000$ particles (averaged over 16 ensembles).

\vskip +0.1in
\noindent{\bf Shear Protocols:}
Subsequently, the system was quenched to $T=0.001$ and sheared with and without activity at a finite temperature ($T$) and finite shear rate ($\dot{\gamma}$) using SLLOD equations in conjunction with Lees-Edwards boundary conditions \cite{10.1063/1.479358, 10.1063/1.1858861, 10.1093/oso/9780198803195.001.0001}. A Gaussian thermostat was used to maintain the temperature constant. By keeping the concentration of activity fixed, we varied the activity parameters $\tau_p$ and $f_0$ to study the mechanical response at different strain rates. After deforming the system, we studied the yielding behavior and shear band formation by measuring the non-affine displacement of the particles using $D^2 _{min}$ \cite{PhysRevE.57.7192}. The cut-off radius for the $D^2 _{min}$ calculation was chosen to be $2.5$. 

\vskip +0.1in
\noindent{\bf Calculation of Volume Fraction:}
The volume fraction is defined as the ratio of the volume or area (in 3d or 2D, respectively) occupied by the shear band to the total volume or total area. To calculate this, we first determine the non-affine displacements of particles by dividing the system into small cells in 2D and evaluating $D^2_{min}$ for each cell. A cell is considered to be part of the shear band region if its local $D^2_{min}$ exceeds a predefined cutoff $0.001$ in 2D (scaled to unit box length) in reduced Lennard-Jones (LJ) units. Using this technique, we replicate a binary image of the shear band snapshot, as shown in Fig.~4 of the supplementary material. The cutoff values are chosen to capture the initial increase in the volume fraction. A similar protocol is applied in 3D, where the system is divided into small cubes instead of squares, and the shear band volume is identified using a $D^2_{min}$ cutoff of $0.01$. \JH{In both 2D and 3D, it is obtained via an average over $16$ and $8$ ensembles for a system sizes of $N=10000$ and $N=100000$, respectively.} 

\section*{Author Contributions}
\SK{SK conceived the project. SK and JH supervised the project. RP performed the research and simulations. RP, JH, and SK designed analysis methods. RP performed all the analyses. RP, JH, and SK wrote the paper jointly.}

\begin{acknowledgments}
\SK{We want to thank Rishabh Sharma, Anoop Mutneja, Maria Ozawa, Srikanth Sastry, and Viswas Venkatesh for many useful discussions. We are grateful to the Alexander von Humboldt Foundation for financial support. S.K. would like to acknowledge the Alexander von Humboldt fellowship for experienced researchers for generous support during S.K.'s visits to Heinrich Heine University of Düsseldorf. We acknowledge funding by intramural funds at TIFR Hyderabad from the Department of Atomic Energy (DAE) under Project Identification No. RTI 4007. SK acknowledges the Swarna Jayanti Fellowship grants DST/SJF/PSA01/2018-19 and SB/SFJ/2019-20/05 from the Science and Engineering Research Board (SERB) and Department of Science and Technology (DST) and the National Super Computing Mission (NSM) grant $\mathrm{DST/NSM/R\&D\_HPC\_Applications/2021/29}$ for generous funding. SK would like to acknowledge the research support from the MATRICES Grant MTR/2023/000079, funded by SERB. Most of the computations are done using the HPC clusters procured using Swarna Jayanti Fellowship grants DST/SJF/PSA01/2018-19, SB/SFJ/2019-20/05 and Core Research Grant CRG/2019/005373.}
\end{acknowledgments}

\bibliography{activeYieldingUltraStable}

\end{document}


\title{\textbf{Supplementary Information - Mitigating Brittle Failure in Ultrastable Glasses via Active Particles Doping}}
\author{Rashmi Priya$^1$} 
\author{J\"urgen Horbach$^2$}
\author{Smarajit Karmakar$^1$} 
 \email{smarajit@tifrh.res.in}
\affiliation{ $^1$ Tata Institute of Fundamental Research, 36/P, Gopanpally Village, Serilingampally Mandal,
Ranga Reddy District, Hyderabad, India 50046\\ $^2$ Institut für Theoretische Physik II - Soft Matter, Heinrich-Heine-Universität, Düsseldorf, Germany}
\maketitle

\section{Sheared Ultrastable Glass in 3D}
\begin{figure*}
\includegraphics[width=0.99\textwidth]{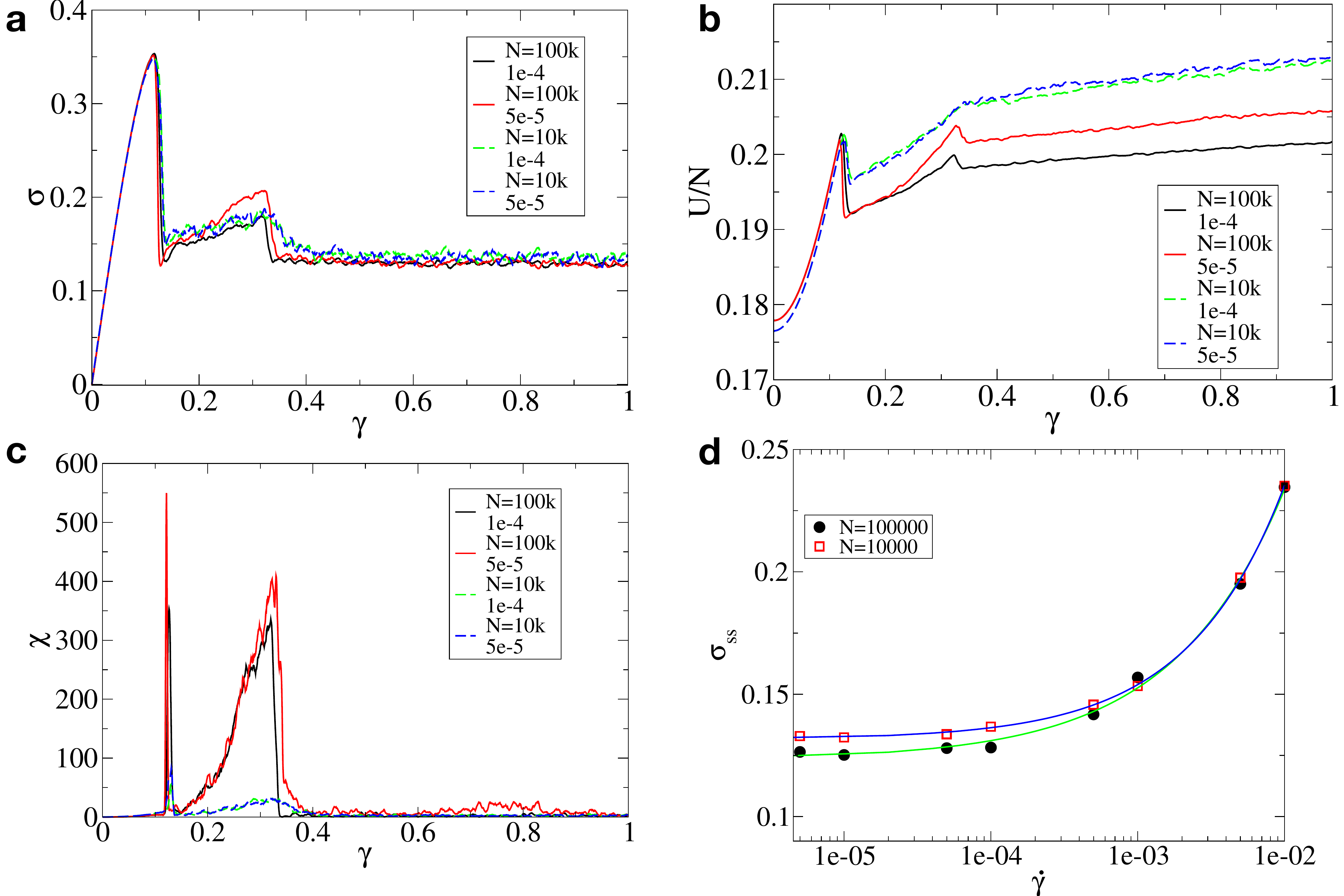}
\caption{{\bf Ultrastable Glassy System in 3D:} Figure (a), (b), and (c) present the stress-strain curve, potential energy per particle, and susceptibility, respectively, for two different system sizes. The dotted line corresponds to \( N = 10k \), and the solid line to \( N = 100k \). In Figure (d), we plot the Herschel-Bulkley curve for both cases, demonstrating shear-thinning behavior consistent with findings from previous studies.}
\label{fig:Passive}
\end{figure*}
In Fig.~\ref{fig:Passive}, we show the mechanical response of an ultrastable glass under shear for two system sizes, $10,000\, (10k)$ and $100,000\, (100k)$, in 3D. Figure (a) shows the stress-strain curves at two different strain rates, $\dot{\gamma} = 1 \times 10^{-4}, 5 \times 10^{-5}$. In (b) and (c), we plot the system's potential per particle and susceptibility. We show that the system exhibits shear-thinning behavior, characterized by the Herschel–Bulkley model with a flow index less than $1$.

\section{Mechanical response of Ultrastable glass in the Presence of self-propelled particles}
\begin{figure*}[htpb]
\includegraphics[width=0.99\textwidth]{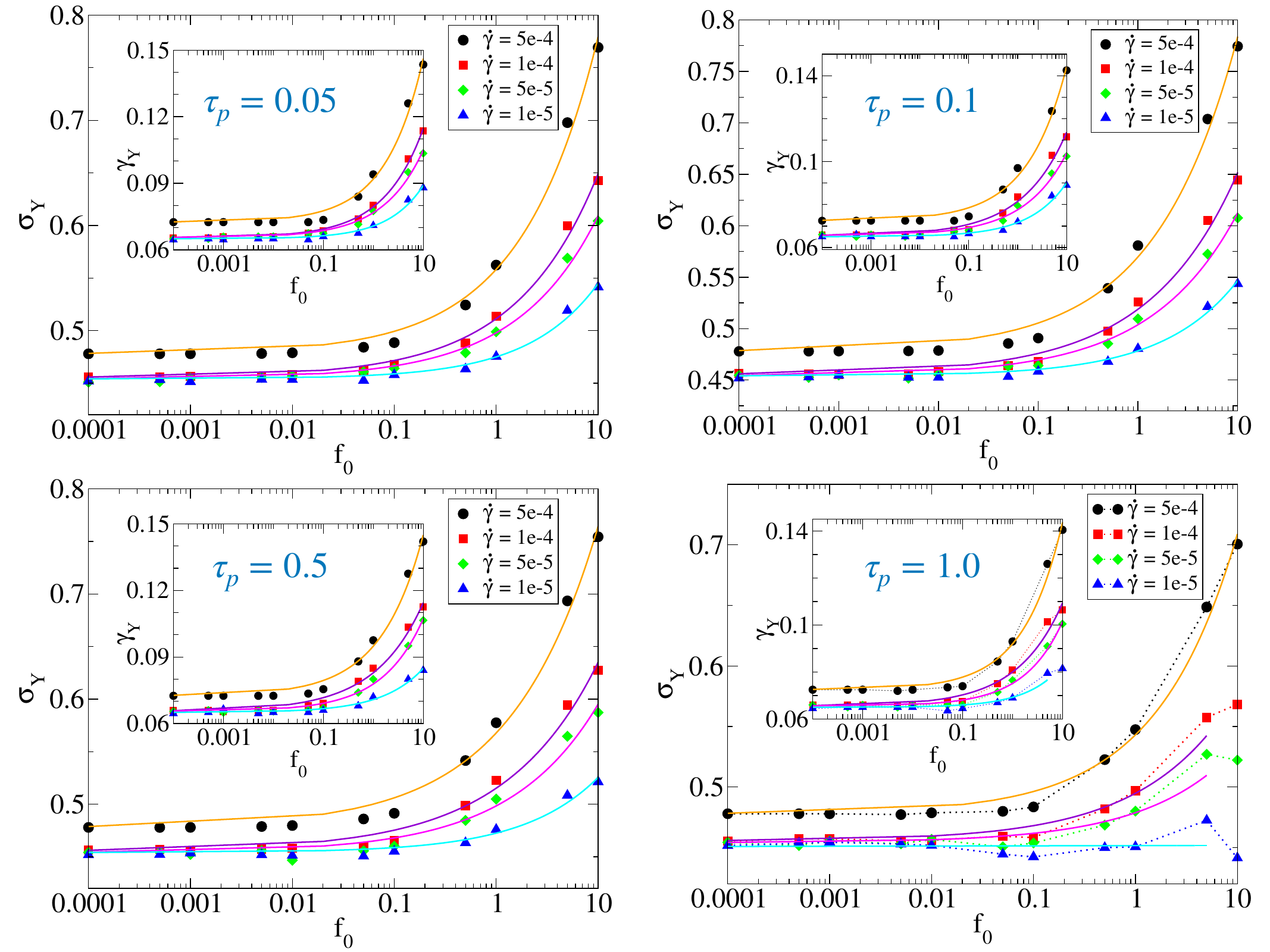}
\caption{{\bf Strain Rate Dependence of Yielding at Different $\tau_p$:} For lower $\tau_p$ values ($0.05, 0.1, 0.5$), the system exhibits shear-thinning behavior with activity across all strain rates studied, as shown in (a), (b), and (c) (for the 2D case). As $\tau_p$ increases to $1.0$, the yield stress at lower strain rates initially rises with increasing $f_0$ in the intermediate regime; however, further increases in $f_0$ cause the yield stress to decrease.}
\label{fig:YieldStress_diffTaup}
\end{figure*}

\begin{figure*}[htpb]
\includegraphics[width=0.99\textwidth]{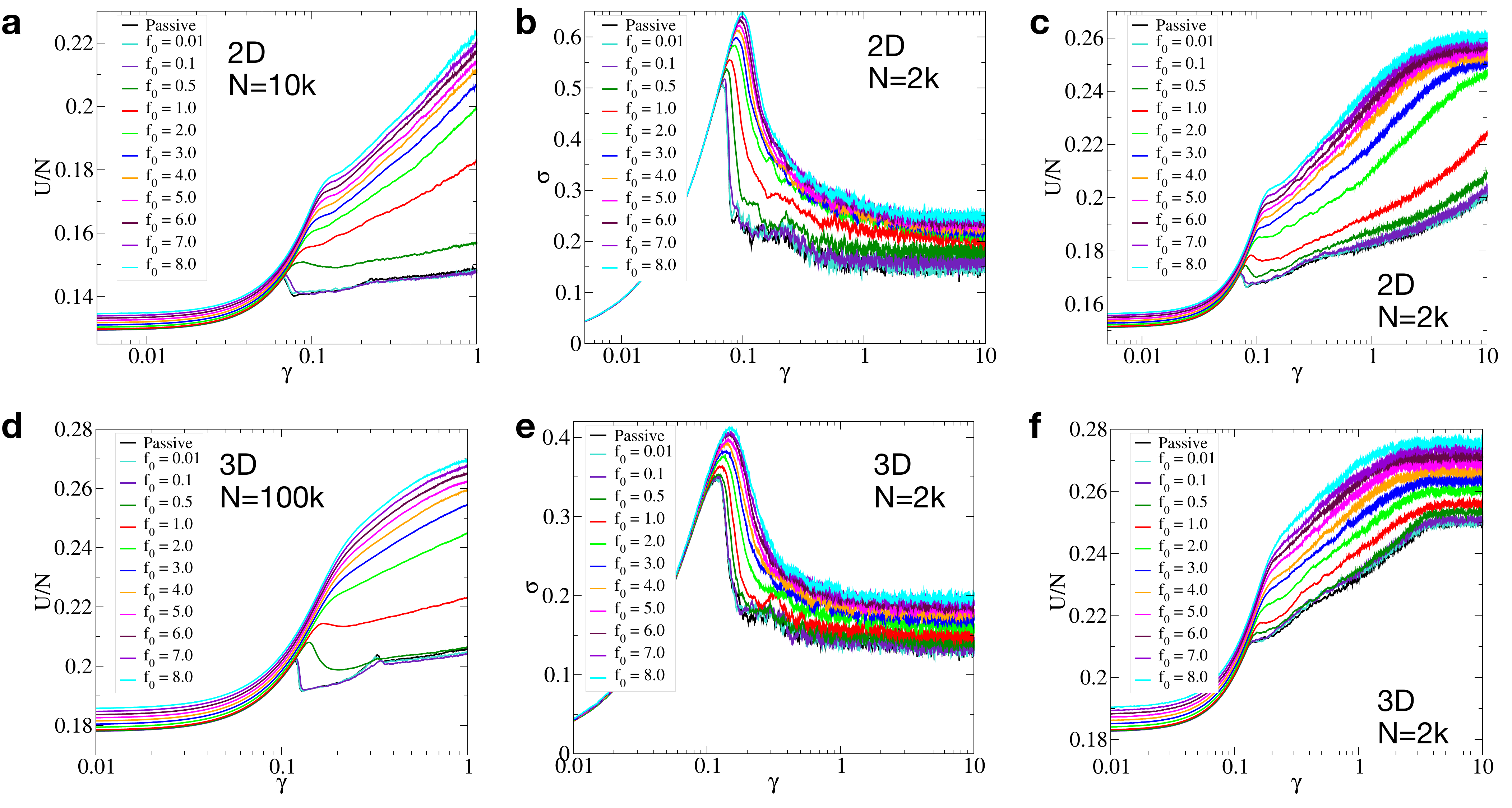}
\caption{{\bf Steady State Analysis in Two and Three Dimensions:} This figure demonstrates that system size plays a crucial role in reaching a steady state. Panels (a) and (d) show that the potential energy for 2D and 3D systems with $N = 10^4$ and $N = 10^5$ continues to increase, indicating that these systems remain far from steady state. Therefore, we examine a smaller system size ($N = 2000$) in both 2D and 3D, and plot the stress and potential energy responses with increasing strain in panels (b) and (c) for 2D, and panels (e) and (f) for 3D. In 2D, we observe that while the stress plateaus, the potential energy continues to increase for lower active forces. However, as $f_0$ increases, the system reaches steady state more rapidly in both dimensions. This acceleration occurs due to the formation of multiple shear bands that span the system more easily when the network structure is homogeneously distributed throughout the system. In contrast, the 3D system clearly achieves a well-defined steady state across all measured quantities and is therefore better suited for studying steady-state behavior in active ultrastable systems.}
\label{fig:SteadyStateAnalysis}
\end{figure*}

We show the shear thinning behavior for different persistence times ($0.05, 0.1, 0.5$) and strain rates ($5 \times 10^{-4}, 1 \times 10^{-4}, 5 \times 10^{-5}, 1 \times 10^{-5}$) in Fig.~\ref{fig:YieldStress_diffTaup}. As persistence time increases to $\tau_P=1.0$, we observe an increase in yield stress followed by a decrease at $\dot{\gamma}= 5 \times 10^{-5}, 1 \times 10^{-5}$, indicating a non-monotonic behavior of yield stress with increasing $f_0$. Nevertheless, shear thinning behavior is still observed at faster strain rates. As $\tau_p$ increases further, the strain rate range over which the enhanced mechanical response is observed decreases. 

The non-monotonic effect with increasing $f_0$ becomes more pronounced at lower strain rates, where slower deformation allows shear band propagation, while faster strain rates inhibit it. Increasing $\tau_p$ further at higher $f_0$ facilitates rearrangements and stress release, leading to a reduction in yield stress. 

After studying the yielding behavior, it becomes essential to investigate the flow characteristics of the system, particularly in the steady-state regime, where transient effects have relaxed and the system’s stress fluctuates around a well-defined mean. To assess whether the system has reached this steady state, we analyze the potential energy, which serves as a sensitive measure of structural relaxation. When the potential energy plateaus, it indicates that the system has stabilized. We therefore plot the potential energy for larger system sizes ($N = 10^4$ in 2D and $N = 10^5$ in 3D) and observe that within the accessible deformation window, the potential energy continues to increase, suggesting that these systems have not yet reached steady state and may require much larger strains to do so. To overcome this limitation, we consider a smaller system size ($N = 2000$) in both 2D and 3D and examine the stress and potential energy response up to a large strain of $\gamma = 10.0$. We find that in 3D, both the stress and potential energy stabilize, indicating a clear steady state. In 2D, while the stress reaches a plateau, the potential energy continues to rise, particularly at lower active forces (\$f\_0\$), including the passive case—likely due to the system-spanning shear bands or less branched network structures. Based on these observations, the 3D system with $N = 2000$ is well-suited for studying the flow curve, as it reliably reaches a steady state within accessible deformation.


\section{Shear Band Evolution with Increasing Activity}

\begin{figure*}[htpb]
\includegraphics[width=0.99\textwidth]{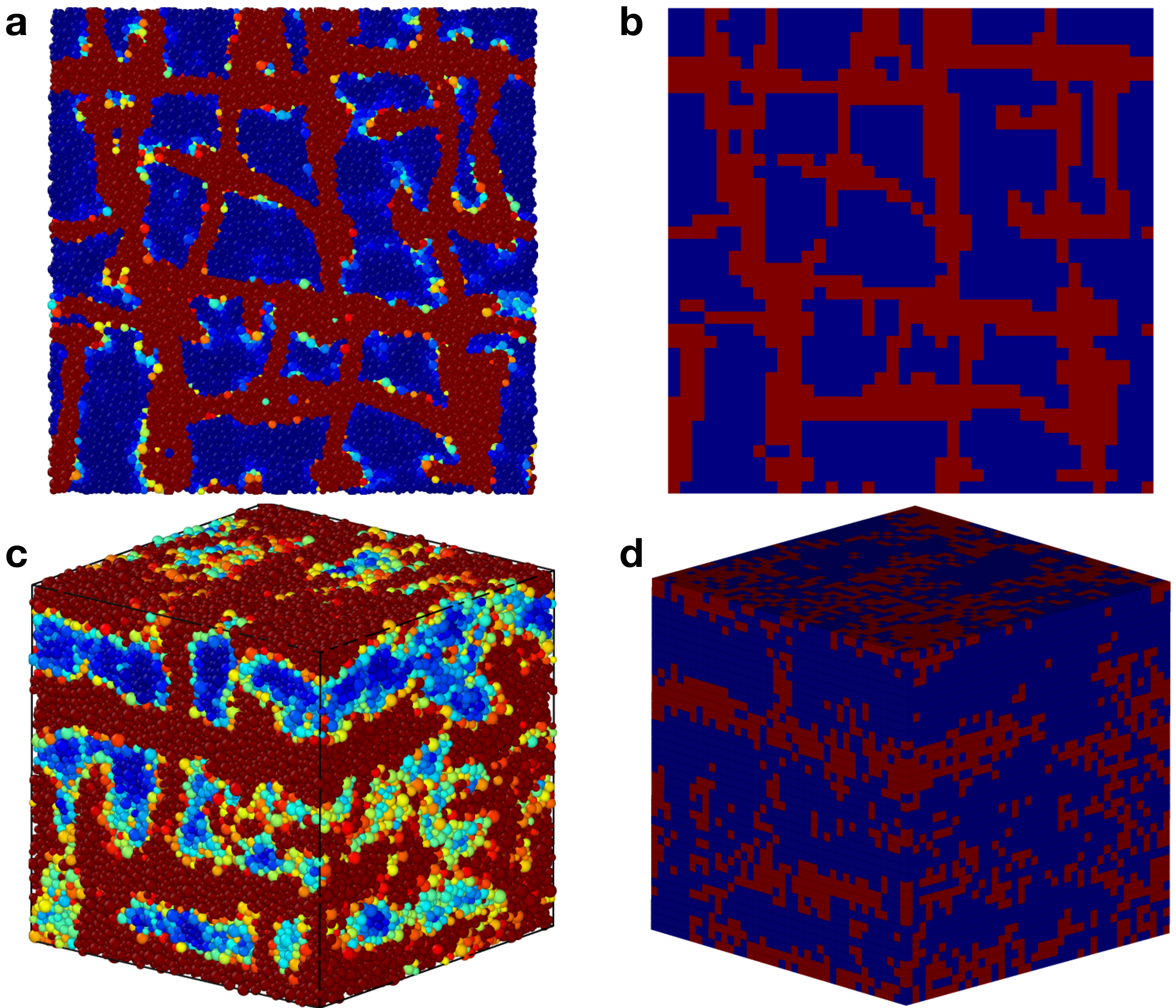}
\caption{{\bf Binary Mapping of System in 2D and 3D}: The images shown in (b) and (d) are binary representations of the actual systems shown in (a) and (c) in 3D and 2D, respectively. The region in red correspond to shear bands, while regions lie outside the shear bands.}
\label{fig:Binaryimage}
\end{figure*}

\begin{figure*}[htpb]
\includegraphics[width=0.99\textwidth]{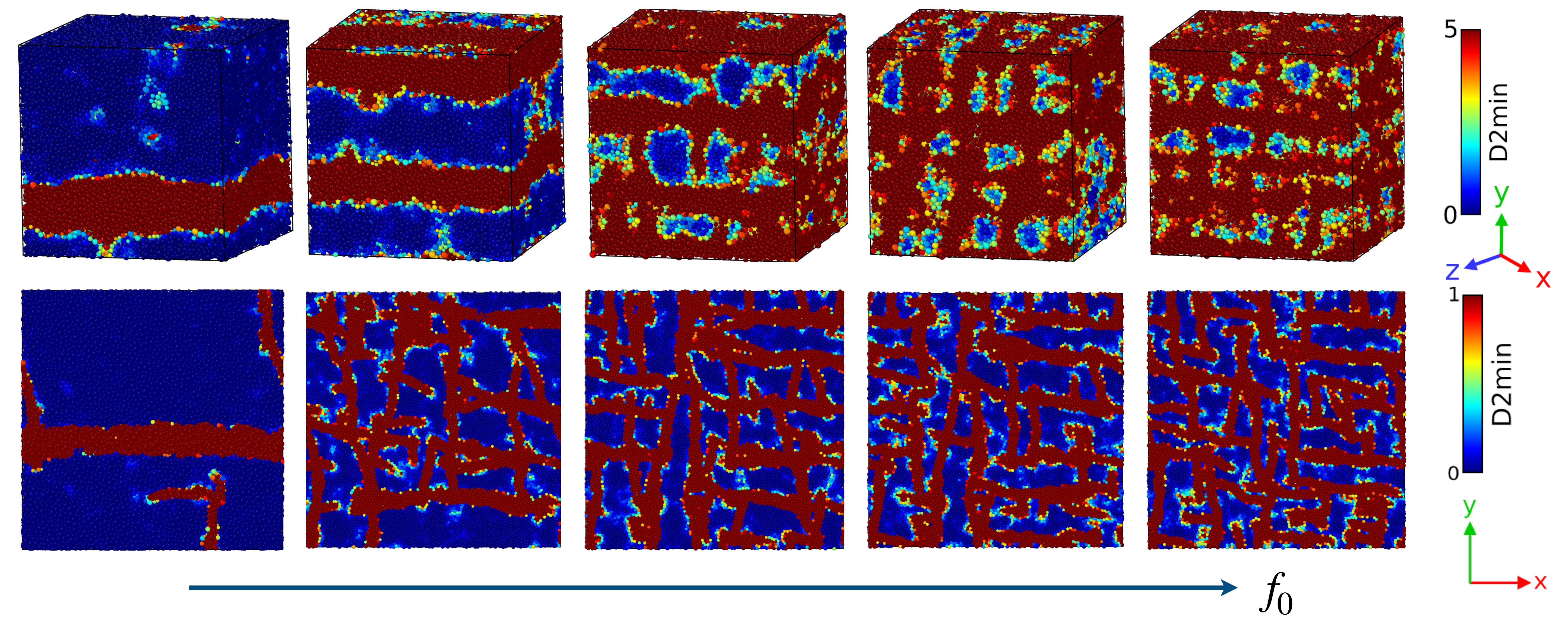}
\caption{{\bf Shear band at $\dot{\gamma} = 1\times 10^{-4}$ and $\tau_p=0.1$}: Evolution of shear band with increasing $f_0$ values $0.0, 1.0, 2.0, 3.0, 4.0$ at $\tau_p=0.1$, strain rate $\dot{\gamma} = 1\times 10^{-4}$ and strain $\gamma = 0.20$ in 2D and 3D.}
\label{fig:SB_1}
\end{figure*}

\begin{figure*}[htpb]
\includegraphics[width=0.99\textwidth]{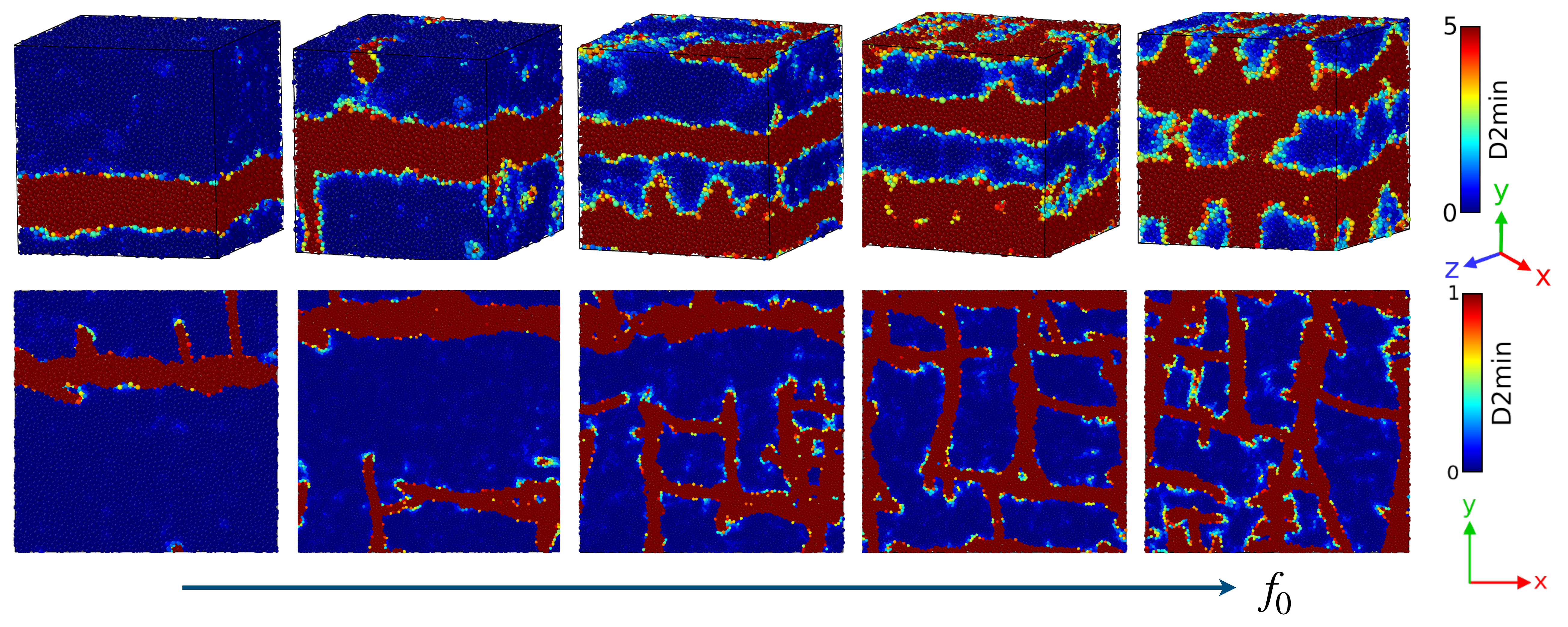}
\caption{{\bf Shear band at $\dot{\gamma} = 1\times 10^{-5}$ and $\tau_p=0.1$}: Evolution of shear band with increasing $f_0$ values $0.0, 1.0, 2.0, 3.0, 4.0$ at $\tau_p=0.1$, strain rate $\dot{\gamma} = 1\times 10^{-5}$ and $\gamma = 0.20$ in 2D and $\gamma = 0.40$ in 3D.}
\label{fig:SB_2}
\end{figure*}

We first show a binary mapping of the system to a replica, which is used for all shear band analyses. In 2D, the system is divided into small cells, and in 3D, it is divided into small cubes. The non-affine displacements of particles within each cell or cube are quantified using $D^2_{min}$. A cell or cube is classified as part of the shear band region (marked in red) if its local $D^2_{min}$ exceeds a predefined cutoff in reduced Lennard-Jones (LJ) units, with the cell size scaled to unit box length. Otherwise, it is considered outside the shear band region as shown in blue. Using this method, we generate a binary image that replicates the shear band regions, as illustrated in Fig.~\ref{fig:Binaryimage} (b) and (d).

We also show shear band evolution at two additional strain rates $\dot{\gamma}=1 \times 10^{-4}, 1 \times 10^{-5}$ at $\tau_p = 0.1$, where we observe that as $f_0$ increases, the network of shear bands becomes denser as shown in Fig.~\ref{fig:SB_1} and Fig.~\ref{fig:SB_2}. At faster strain rates, $\dot{\gamma} = 1\times 10^{-4}$, a smaller self-propulsion force, $f_0=1.0$ is sufficient to produce multiple shear bands, whereas at slower deformation rate, $\dot{\gamma} = 1\times 10^{-5}$, a higher value of active force, $f_0 \geq 8.0$ is required to form similar networks.

\section{Scaling of Yield Stress with Strain Rate and Active Force}

\begin{figure*}[htpb]
\includegraphics[width=0.7\textwidth]{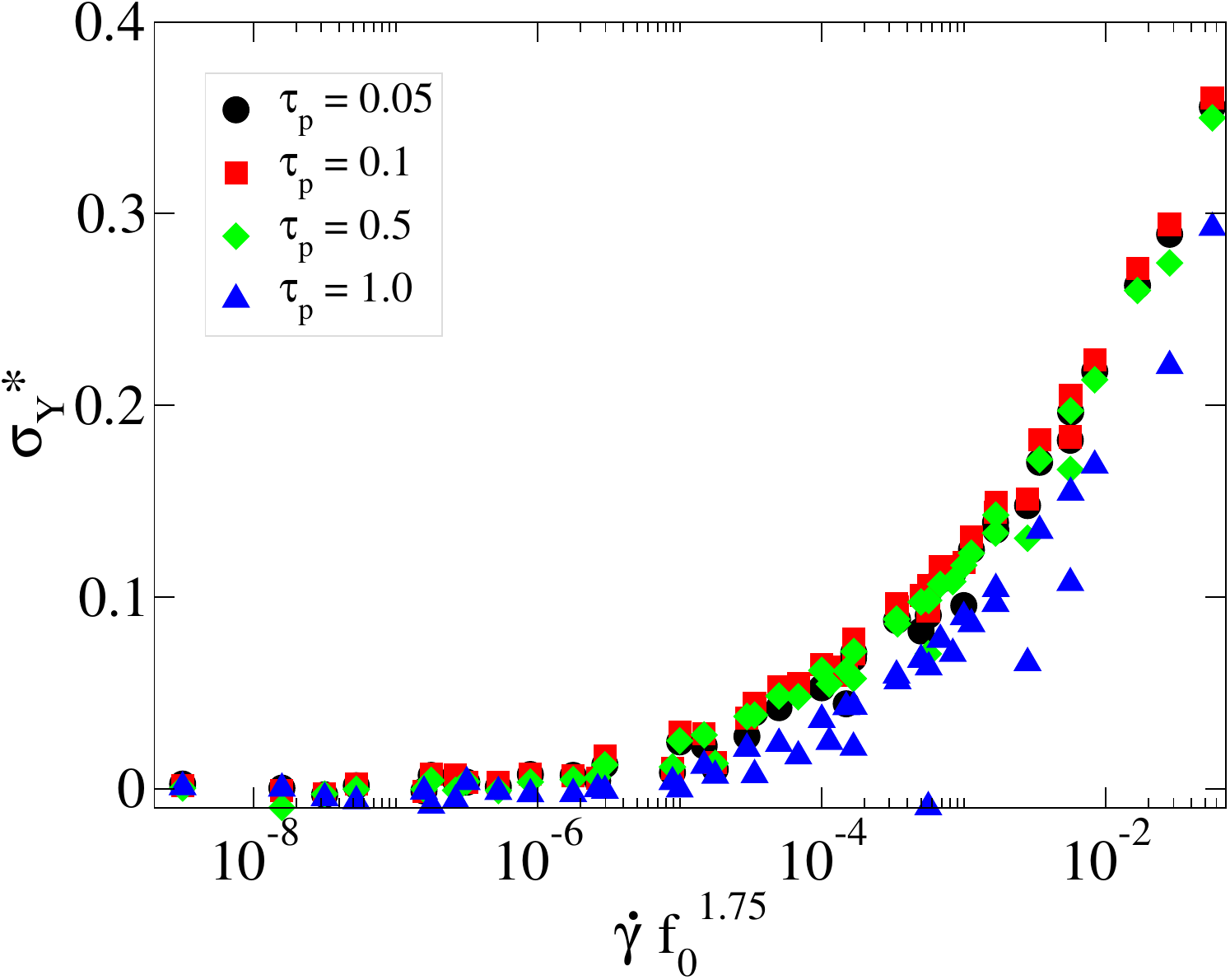}
\caption{{\bf Yield Stress Collapse Across Different Persistence Times:} We demonstrate that the yield stress data for all values of $\tau_p = 0.05, 0.1, 0.5, 1.0$ exhibit a collapse onto a single curve for each $\tau_p$ when the x-axis is rescaled as $\dot{\gamma} f_0^{1.75}$ (in 2D). For $\tau_p=1.0$, the data start deviating from collapse because a low strain rate and high $f_0$ lead to a non-monotonic increase followed by a decrease in yield stress, as observed in Fig.~\ref{fig:YieldStress_diffTaup}.} 
\label{fig:Data_collapsed_diffTaup}
\end{figure*}

In Fig.~\ref{fig:Data_collapsed_diffTaup}, we show the collapse of yield stress data presented in Fig.~\ref{fig:YieldStress_diffTaup} with increasing $f_0$, at different strain rates for different persistence times in 2D. We observe a strong collapse for $\tau_p = 0.05, 0.1, 0.5$ when the x-axis is rescaled to $\dot{\gamma} f_0^{1.75}$. The y-axis is modified yield stress $\sigma_Y^*$ given by:
\begin{eqnarray}
\sigma_Y^* = \sigma_Y(\dot\gamma, f_0) - \sigma_{AQS}^P - \alpha_0 \dot{\gamma}^{\beta_0}
\label{eq1}
\end{eqnarray}
where $\sigma_{AQS}^P$, $\alpha_0$ and $\beta_0$ are deduced from Fig~\ref{fig:Passive}. 

\begin{eqnarray}
\sigma_Y^* = {\alpha} ({\dot{\gamma}f_0^m}) ^ {\beta}
\label{eq2}
\end{eqnarray}

The data follows the eq.~\ref{eq2}, where $\alpha=1.0$ and $\beta < 1$, which can be expressed in its final form as:
\begin{eqnarray}
\sigma_Y(\dot\gamma, f_0) =  \sigma_{AQS}^P +  \alpha_0 \dot{\gamma}^{\beta_0} + {\alpha} ({\dot{\gamma}f_0^m}) ^ {\beta},
\label{eq3}
\end{eqnarray}

However, as $\tau_p$ increases to $1.0$, the collapse begins to deviate slightly, since at lower strain rates, increasing activity decreases yield stress, as discussed above. The observed collapse reinforces that the interplay between $\dot{\gamma}$ and $f_0$ governs the yield stress behavior across different persistence times.

\section{Yielding Behavior Across Ensembles in Isomorphic Systems}

\begin{figure*}[htpb]
\includegraphics[width=0.7\textwidth]{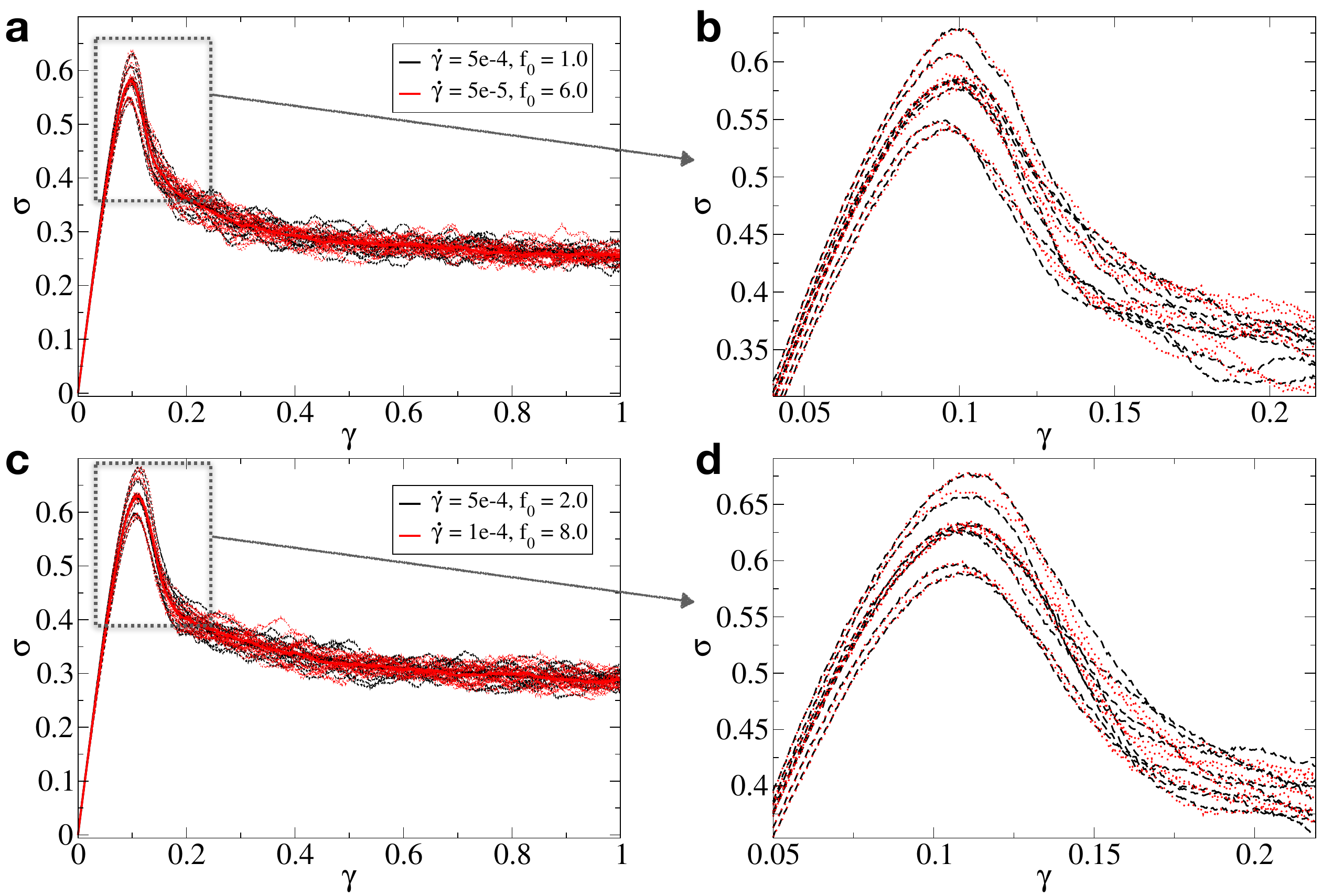}
\caption{{\bf Stress-Strain Response for Two Isomorphs:} The stress-strain curve for two isomorphic curves is plotted for all ensembles (in 2D for $N = 10000$), demonstrating similar yielding behavior before and after the yield point. As the system approaches a steady state, the stress values fluctuate around the mean.}
\label{fig:Isomorph_within_Ensembles}
\end{figure*}

\begin{figure*}[htpb]
\includegraphics[width=0.99\textwidth]{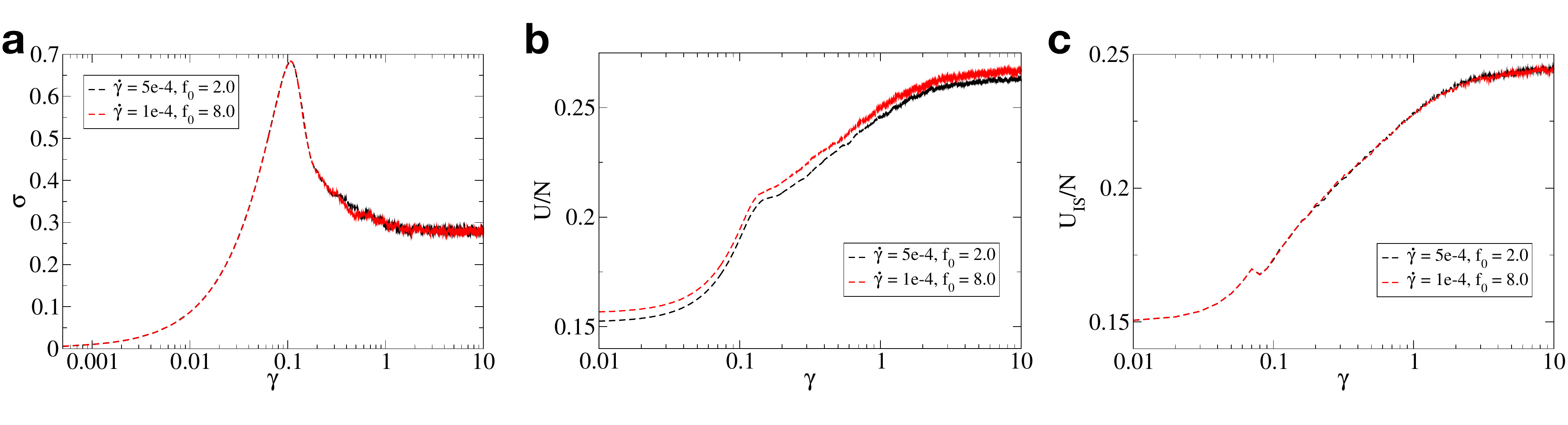}
\caption{{\bf Isomorphs in steady-state:} In this figure, we show the steady-state behavior of a system with size $N = 2000$ in 2D. Under larger deformation, the system's stress, potential energy, and inherent state energy reach a steady state. We demonstrate that, in the steady state, the stress and inherent state energy remain the same for two different combinations of $\dot{\gamma}$ and $f_0$, whereas the potential energy differs.}
\label{fig:Isomorph_N2k_Steadystatebehavior}
\end{figure*}

\begin{figure*}[htpb]
\includegraphics[width=0.99\textwidth]{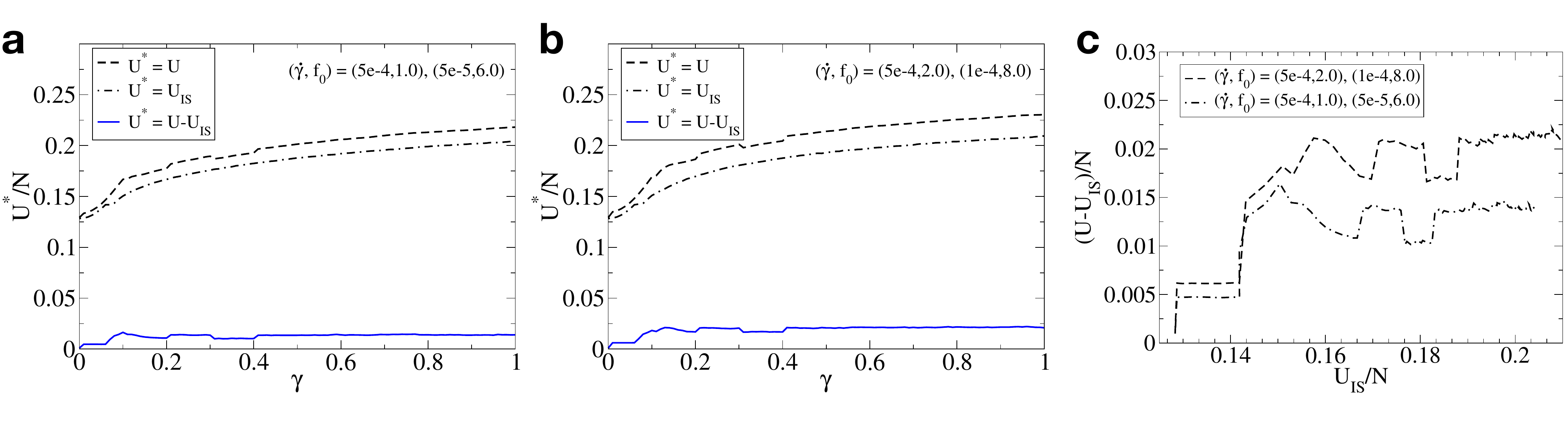}
\caption{In this figure, we quantify the difference between the potential energy and the inherent‑state energy for two isomorphic cases (in 2D for $N = 10000$): $(\dot{\gamma}, f_0) = (5\times10^{-4}, 1.0), (5\times10^{-5}, 6.0)$ and $(\dot{\gamma}, f_0) = (5\times10^{-4}, 2.0), (1\times10^{-4}, 8.0)$. Panels (a) and (b) show the potential energy, the inherent‑state energy, and their difference ($U^*=(U-U_{IS})$) while switching at intervals of $0.1$, $0.2$, $0.3$, and $0.4$ from one pair to the other for both cases. The difference follows a pattern similar to that of the potential energy, and when this difference is plotted against the inherent‑state energy, $(U-U_{IS})/N$ jumps to a new plateau and then remains constant within that state until the next switch.}
\label{fig:Isomorph_diffPE_InherentstateEnergy}
\end{figure*}

We present the stress-strain curve for different ensembles corresponding to two distinct pairs of isomorphs.
\begin{itemize}
    \item ($\dot{\gamma} = 5 \times 10^{-4}$, $f_0=1.0$),( $\dot{\gamma} = 5 \times 10^{-5}$, $f_0=6.0$)
    \item ($\dot{\gamma} = 5 \times 10^{-4}$, $f_0=2.0$),( $\dot{\gamma} = 1 \times 10^{-4}$, $f_0=8.0$).
\end{itemize}

The results show that the yielding behavior of an ensemble for a pair of isomorphs is exactly the same. However, beyond the yield point, the responses start to deviate due to the formation of multiple shear band networks. This deviation occurs because both the polydispersity of the system and the random selection of 20\% active particles lead to different spatial distributions of active particles between isomorphs.

We also show that the behavior remains the same even when the system reaches steady state. For this, we consider a smaller system size $N=2000$ in 2D, where the system reaches steady state under larger deformation $\gamma = 10.0$. We examine the evolution of stress, potential energy and inherent state energy responses of the isomorphs: $\dot{\gamma} = 5 \times 10^{-4}$, $f_0=2.0$),( $\dot{\gamma} = 1 \times 10^{-4}$, $f_0=8.0$ with strain, as shown in Fig.~\ref{fig:Isomorph_N2k_Steadystatebehavior}.

We also show the difference in potential energy and inherent state energy in Fig.~\ref{fig:Isomorph_diffPE_InherentstateEnergy} for $N=10000$ in 2D, and observe that when switching from one combination of $(\dot{\gamma}, f_0) = (5\times10^{-4}, 1.0), (5\times10^{-5}, 6.0)$ to another $(\dot{\gamma}, f_0) = (5\times10^{-4}, 2.0), (1\times10^{-4}, 8.0)$, the difference exhibits a jump at the switching point and remains constant otherwise.

\section{Non-Monotonic Mechanical Response}

\begin{figure*}[htpb]
\includegraphics[width=0.99\textwidth]{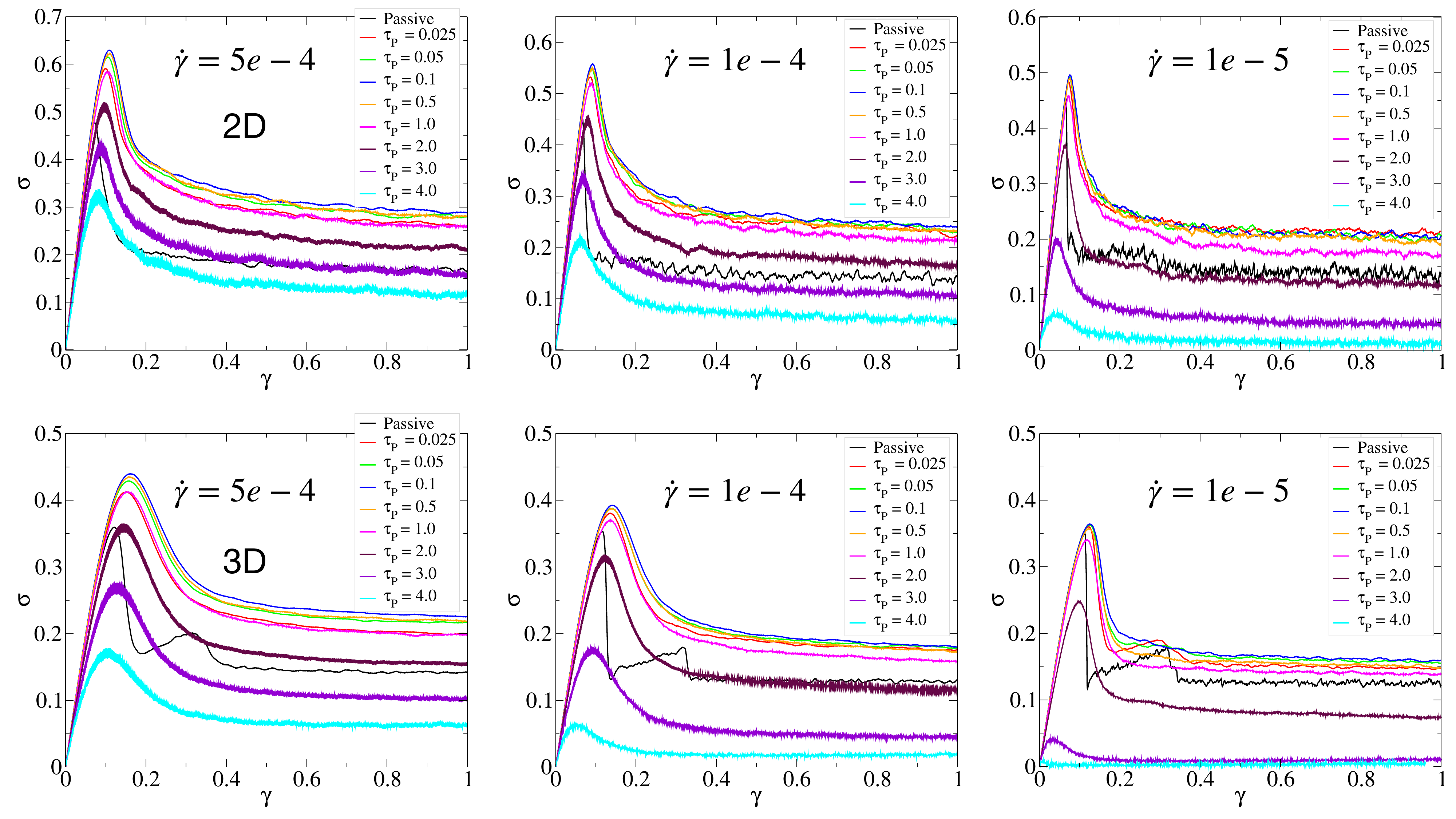}
\caption{{\bf Non-Monotonic Yielding with $\tau_p$ at Different Strain Rates:} The stress-strain curve for $f_0 = 2.0$ is shown at a strain rates: $\dot{\gamma} = 5\times10^{-4}, 1\times 10^{-4}, 1\times 10^{-5}$. The top panel displays the results for 2D, while the bottom panel shows the results for 3D. This figure confirms that the non-monotonicity behavior is independent of the strain rate and is consistently present in the system. However, the value of $\tau_p$ at which the stress overshoot transitions to an undershoot varies; a larger strain rate corresponds to a larger $\tau_p$ value up to which the stress overshoot is observed.}
\label{fig:StressStrainatf2}
\end{figure*}

\begin{figure*}[htpb]
\includegraphics[width=0.99\textwidth]{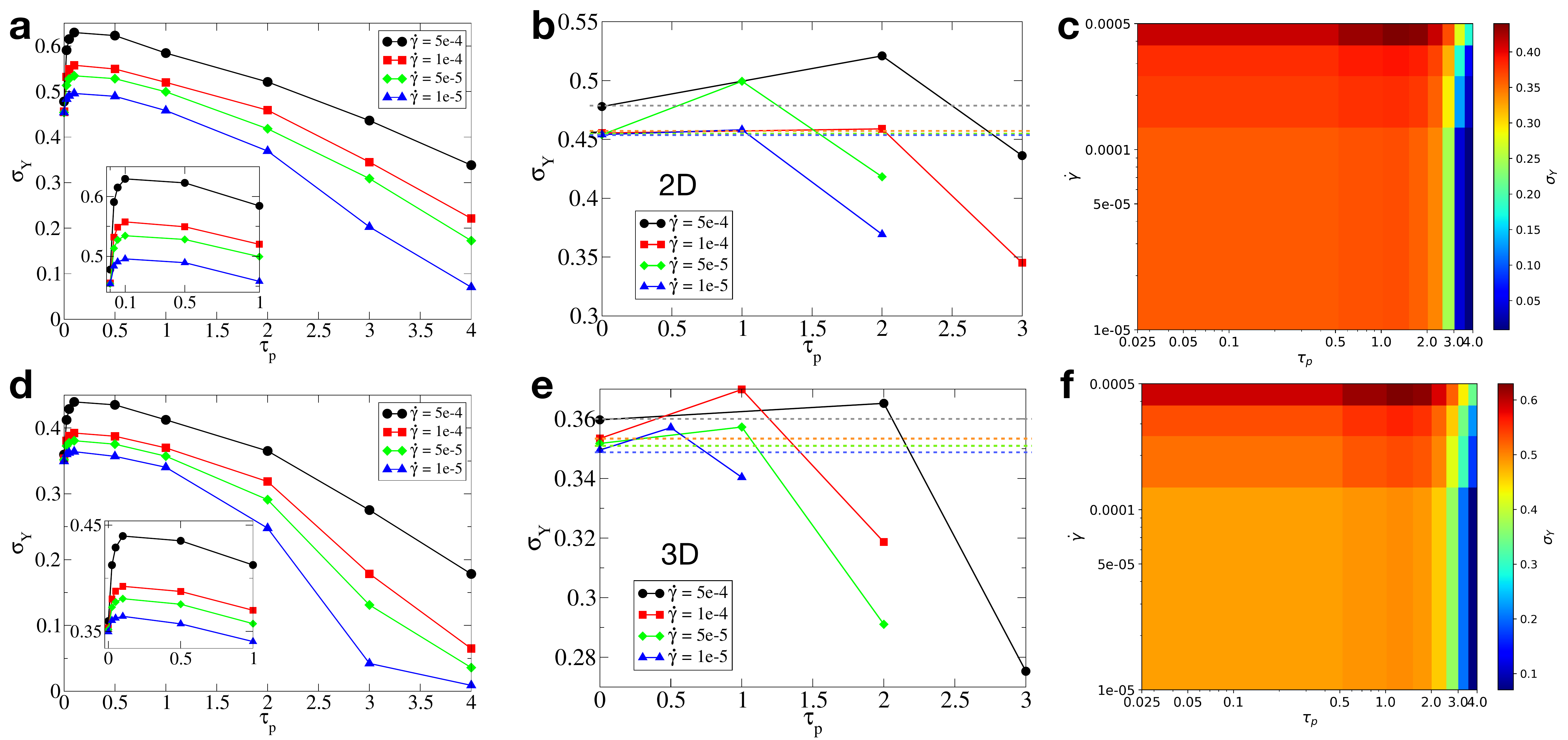}
\caption{{\bf Yield Stress for Different $\tau_p$:} The panels (a) and (d) present the yield stress ($\sigma_Y$) as a function of the persistence time $\tau_p$ for different strain rates ($\dot{\gamma} = 5 \times 10^{-4}, 1 \times 10^{-4}, 5 \times 10^{-5}, 1 \times 10^{-5}$) in 2D and 3D systems respectively. Panels (b) and (e) present zoomed-in views of (a) and (d), focusing specifically on the data points where the yield stress is slightly above and slightly below the reference passive yield stress value. The reference passive yield stress is indicated by the dotted line at various strain rates. These panels highlight that the $\tau_p$ value, up to which an increase in yield stress is observed, increases with higher strain rates. Panels (c) and (f) present heatmaps of the yield stress as a function of $\tau_p$ for increasing $\dot{\gamma}$, and we see the decrease in yield stress as $\tau_p$ approaches $>2.0$ in 2D and 3D, respectively. These plots illustrate how the relationship between $\tau_p$, strain rate, and yield stress transitions between 2D and 3D systems.}
\label{fig:Heatmap}
\end{figure*}

\begin{figure}[htpb]
\includegraphics[width=0.99\textwidth]{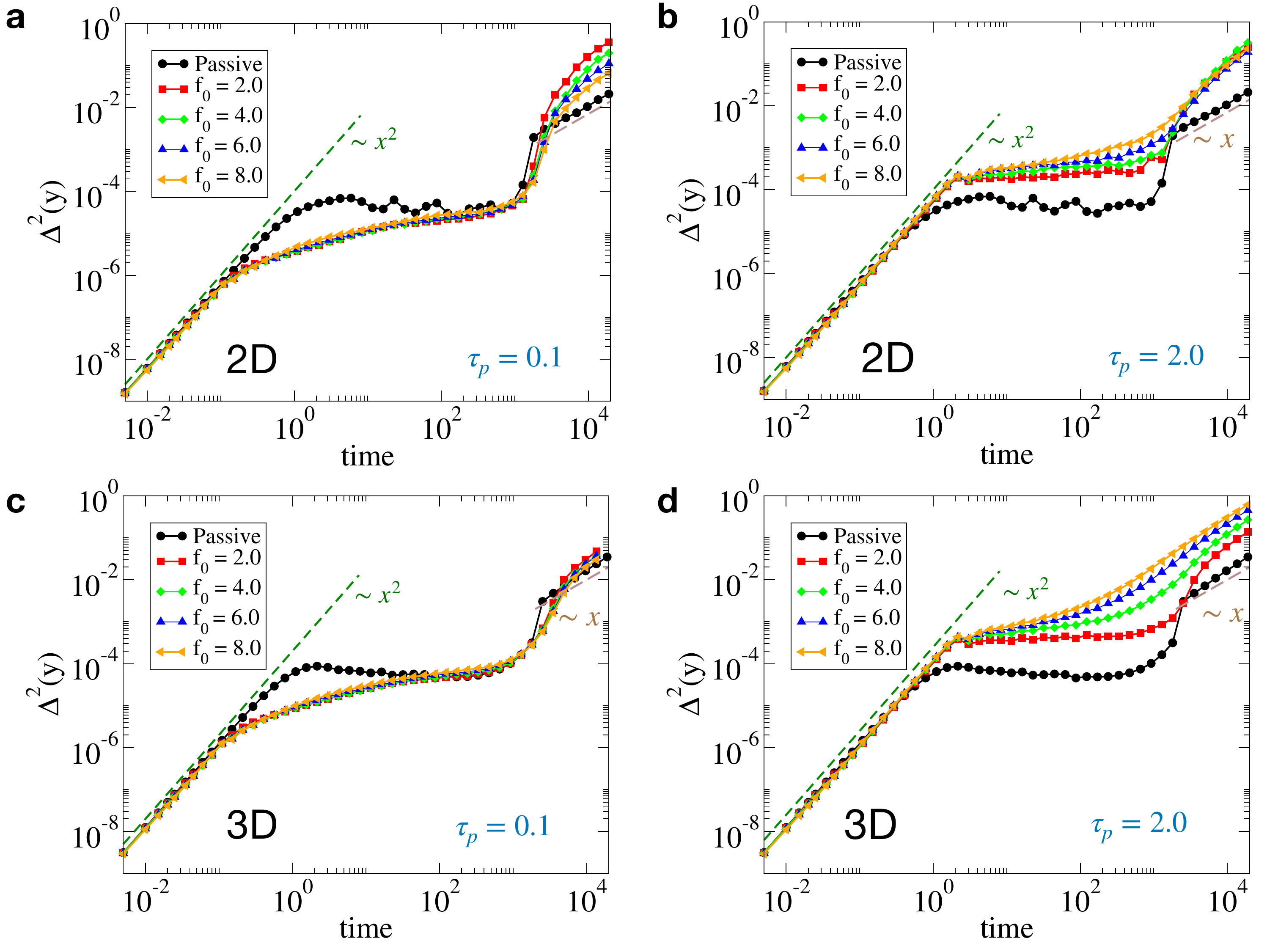}
\caption{{\bf Mean Square Displacement (MSD):} This figure shows the mean square displacement (MSD) of particles in the shear perpendicular direction for 2D (top panel) and 3D (bottom panel) systems at a strain rate of $\dot{\gamma} = 5 \times 10^{-5}$, $f_0$ at different $\tau_p$ values ($0.1, 2.0$). For $\tau_p = 0.1$, the effect of increasing $f_0$ is noticeable only in the diffusive regime, as there is no prior cage-breaking event. However, for $\tau_p = 2.0$, cage-breaking occurs earlier, and the impact of increasing  $f_0$ is evident sooner. Particles transition directly from the ballistic to the diffusive regime, bypassing the cage with increasing $f_0$.}
\label{fig:MSD}
\end{figure}


We observe a non-monotonic behavior in the yield stress of the ultrastable glasses under deformation in the presence of self-propelled particles. In Fig.~\ref{fig:StressStrainatf2}, we show stress-strain curves for different strain rates, $\dot\gamma = 5 \times 10^{-4}, 1 \times 10^{-4}, 1 \times 10^{-5}$, highlighting that the non-monotonic behavior is consistently observed at $\tau_p = 0.1$ across all cases. However, with increasing strain rate, there is a shift to a higher value in both the persistence time at which yield stress overshoots the passive case and the persistence time at which $\sigma_Y \rightarrow 0$. 

To further illustrate this, we plot yield stress as a function of $\tau_p$ in Fig.~\ref{fig:Heatmap}, shown in both 2D (top panel) and 3D (bottom panel). In (b) and (e), only three selected data points are shown, along with a dotted line extending from the first data point, which corresponds to the passive system for a given strain rate. $\tau_p=0$, refers to the passive system, while the middle point corresponds to the maximum observed increase in yield stress and the last point marks the first instance where it decreases relative to the passive system. The data shows that curves at faster strain rates (black) correspond to larger $\tau_p$ values, while curves at slower strain rates (blue) correspond to lower $\tau_p$ values. The heatmap reveals a decrease in yield stress as the color transitions from red, emphasizing that the correct combination of strain rates, $\tau_p$ and $f_0$ is crucial for enhancing the mechanical response of the system.

The mean square displacement (MSD) of the system, when plotted for different $f_0$ values at two distinct $\tau_p$ values $0.1, 2.0$, shows differences only in the diffusive regime as seen in Fig.~\ref{fig:MSD}. However, at a larger persistence time ($\tau_p = 2.0$), the system transitions from ballistic to diffusive smoothly, indicating that the particles, due to increased mobility, can escape cages.

\section{Evolution of Spacing Between Shear Bands}

\begin{figure*}[htpb]
\includegraphics[width=0.95\textwidth]{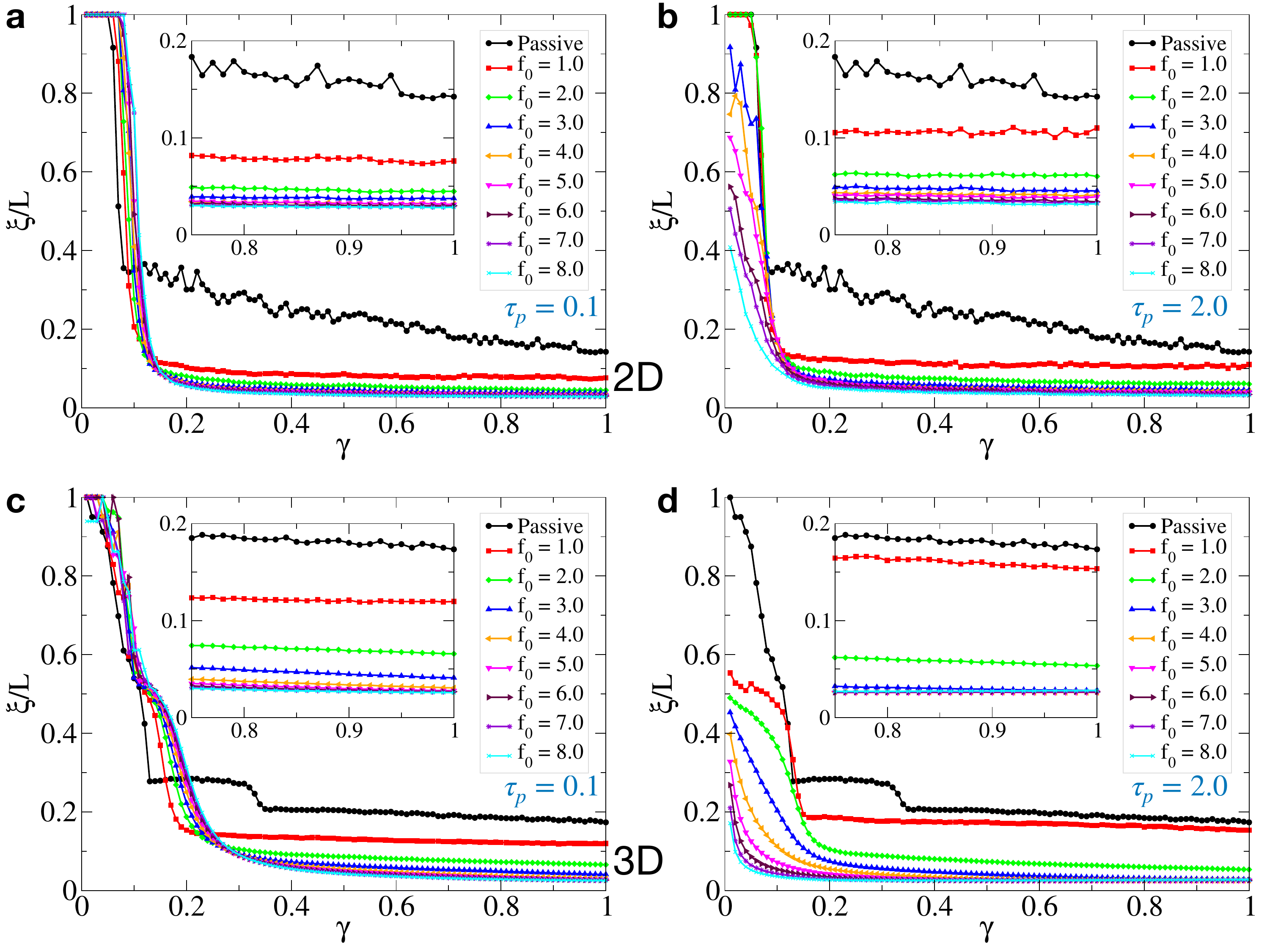}
\caption{{\bf Evolution of Distance Between Shear Bands:}  This figure illustrates the evolution of the separation between shear bands for different $\tau_p$ values $0.1, 2.0$ in 2D (top) and 3D (bottom) systems. The results show that the distance decreases over time, indicating that the shear bands grow and expand as time progresses.}
\label{fig:Distance_bw_SB}
\end{figure*}

We show the evolution of separation between shear band networks with increasing strain at different $\tau_p$ values:
\begin{itemize}
    \item (a) and (c) for $\tau_p = 0.1$ in 2D and 3D, respectively
    \item (b) and (d) for $\tau_p = 2.0$ in 2D and 3D, respectively
\end{itemize}

To find the separation between two shear bands we replicate the binary image of the system in 2D as shown in Fig.~\ref{fig:Binaryimage}. Many lines are drawn in the horizontal and vertical directions. Along each line, the separation length between shear bands is measured by first identifying shear band regions where $D^2_{min}$ exceeds the cutoff. Whenever $D^2_{min}$ falls below the cutoff, we mark that as the end of a shear band region and continue to measure the separation length from this point until $D^2_{min}$ exceeds the cutoff again, indicating entry into the next shear band. Referring to Fig.~\ref{fig:Binaryimage}, the length of the blue regions along the line is measured. The first moment of this length $\xi$, is then averaged over all ensembles. 

In Fig.~\ref{fig:Distance_bw_SB}, we show the $\xi/L$ with increasing strain, where $L=100$.
Initially, the separation starts at $1.0$, as no shear bands are present. It then drops instantly in the passive case as shear bands form, while in active cases, the decrease is more gradual. In the passive system, the separation continues to decrease as the shear bands grow. In active cases, it also decreases initially but eventually saturates due to system size constraints, when the shear band network spans most of the simulation box.

Additionally, we observe that with increasing $f_0$, the separation between shear bands near yielding and in the transient steady-state decreases. As the shear band network becomes denser, the separation between shear bands $\xi/L \rightarrow 0$. However, at $\tau_p = 2.0$ and larger $f_0$, the system does not form shear bands but rather shows mobile regions homogeneously distributed throughout the system. Consequently, $\xi/L \rightarrow 0$ at smaller deformation, which is clearly visible in 3D.